\documentclass[11pt]{article} 

\usepackage[numbers,square, sort&compress]{natbib}
\usepackage{graphicx}
\usepackage{latexsym}
\usepackage[left=.75in,top=.75in,right=.75in,bottom=.75in]{geometry}
\usepackage{amsmath}
\usepackage{booktabs}

\newcommand{\openr}{\hbox{${\rm I\kern-.2em R}$}}
\newcommand{\openn}{\hbox{${\rm I\kern-.2em N}$}}

\newcommand\independent{\protect\mathpalette{\protect\independenT}{\perp}}
    \def\independenT#1#2{\mathrel{\setbox0\hbox{$#1#2$}%
    \copy0\kern-\wd0\mkern4mu\box0}}

\usepackage{caption}

\usepackage{color}
\usepackage{amssymb}
\usepackage[normalem]{ulem}
\usepackage{amsmath}
\newcommand{\stkout}[1]{\ifmmode\text{\sout{\ensuremath{#1}}}\else\sout{#1}\fi}

\usepackage{url}
\usepackage{setspace}

\usepackage{todonotes}

\title{Two-Stage TMLE to Reduce Bias and Improve Efficiency \\ in Cluster Randomized Trials}

\author{Laura B. Balzer, Mark van der Laan, James Ayieko, Moses Kamya, \\ Gabriel Chamie,  Joshua Schwab,
 Diane V. Havlir,  Maya L. Petersen}

\date{\today}

\begin{document}

\maketitle

Cluster randomized trials (CRTs) randomly assign an intervention to groups of individuals (e.g., clinics or communities) and measure outcomes on individuals in those groups. While offering many advantages, this experimental design introduces challenges that are only partially addressed by existing analytic approaches. First, outcomes are often missing for some individuals within clusters. Failing to appropriately adjust for differential outcome measurement can result in biased  estimates and  inference. Second, CRTs often randomize limited numbers of clusters, resulting in chance imbalances on baseline outcome predictors between arms. Failing to adaptively adjust for these imbalances and other predictive covariates can result in efficiency losses. To address these methodological gaps, we propose and evaluate a novel two-stage targeted minimum loss-based estimator (TMLE) to adjust for baseline covariates in a manner that optimizes precision, after controlling for baseline and post-baseline causes of missing outcomes. Finite sample simulations illustrate that our approach can nearly eliminate bias due to differential outcome measurement, while existing CRT estimators yield misleading results and inferences. Application to real data from  the SEARCH community randomized trial demonstrates the gains in efficiency afforded through adaptive adjustment for baseline covariates, after controlling for missingness on individual-level outcomes.\\
 

\section{Introduction}
\label{sec1}

In many trials, treatments are randomly allocated to groups of individuals, such as hospitals, schools, or communities, and outcomes are measured on individuals in those groups. These studies are known as  group or cluster randomized trials (CRTs). They are implemented when the treatment is naturally delivered to the group or when substantial dependence between individuals within groups is expected \citep{HayesMoulton2009, Crespi2016, Turner2017Design, Turner2017Analysis, Murray2020}. 
% CUTTING FOR SPACE
%Such dependence may arise from shared group-level factors and interactions between participants, including the spread of social behaviors and infectious diseases. Indeed, many group-level interventions may benefit both the individuals who receive the intervention and others who do not; CRTs provide an opportunity to assess such spillover effects \citep{HayesMoulton2009,Halloran1991, Halloran1995, Hudgens2008}. Many CRTs are pragmatic in that their goal is to expand results from prior efficacy studies into real-world effectiveness \citep{Platt2010}. 
CRTs are  rapidly increasing in popularity; a recent review found a 280-fold increase in their use from 1995 to 2015 \citep{Murray2018}.  Nonetheless, despite extensive research dedicated to their design and conduct, this review also concluded only half of CRTs were analyzed appropriately. %  \citep{Murray2018}. 
CRTs can provide gold-standard evidence of causality, but they face several methodological challenges.

First, missing participant outcomes occur in over 90\% of CRTs \citep{Fiero2016}. When participants with missing outcomes differ meaningfully from those with measured outcomes, complete-case analyses yield biased estimates \citep{Rubin1976, Robins1995}. 
% Gail1996, Little&Rubin02
This potential for bias is exacerbated when, as commonly occurs, the cluster randomized intervention, itself, influences outcome measurement. Suppose, for example, that the cluster-level intervention increases care engagement, which in turn improves both participants'  outcomes and their chances of having that outcome measured  (Figure~\ref{Fig:DAG} in Supplementary Materials). Here, an unadjusted comparison of outcomes between arms can overestimate or underestimate the treatment effect. Even if key determinants of missingness, such as care engagement, are measured, standard analytic approaches to CRTs will fail to control for this bias, because care engagement simultaneously mediates the treatment-outcome relationship and confounds the missingness-outcome relationship \citep{Robins1986, Robins1995, Robins2009}. 
%Robins2000 *****

Second, CRTs often randomize limited numbers of groups; % \citep{HayesMoulton2009, Crespi2016, Turner2017Design, Turner2017Analysis, Murray2018, Murray2020}. 
a review found a median of 33 clusters randomized \citep{ Selvaraj2013}.  Even when some form of restricted randomization (e.g., pair-matching) is used, CRTs with few clusters are likely to suffer from chance imbalances between treatment arms on baseline determinants of the outcome. Adjustment for these covariates and others predictive of the outcome (hereafter called ``covariate imbalance'') can increase statistical power (e.g., \cite{Fisher1932, Gail1996, MarkBook, HayesMoulton2009,  Colantuoni2015, Turner2017Analysis, Murray2018, Murray2020}).
The adjustment approach is often with an outcome regression, characterizing the expected outcome given the treatment assignment and covariates. This regression   must be  \emph{a priori}-specified to avoid inflating Type-I error rates.
% Pocock2002, Olken2015, Tsiatis2008
Additionally to avoid over-fitting, a limited number of adjustment variables must be selected from a typically large set of candidates, risking forced adjustment for variables that prove useless for, or even detrimental to, precision
\citep{Stephens2013,Balzer2016DataAdapt, Kahan2016}.
Thus, we wish to define a fully pre-specified procedure for CRT analysis that  optimizes statistical power through data-adaptive adjustment  
of baseline covariates, while rigorously preserving Type-I error control.

%%%%%%%%%%%%%%%%%%
% Throughout,  we use ``cluster'', ``group'', and ``community'' interchangeably.
 In this manuscript, we propose and evaluate a novel estimator that  addresses the dual challenges of bias due to missing outcomes and imprecision due to few randomized units in CRTs.
Our approach uses targeted minimum loss-based estimation (TMLE; \cite{MarkBook}) in two stages: first at the individual-level to adjust for differential measurement of individual-level outcomes and second at the cluster-level to improve efficiency when estimating the intervention effect. Therefore,  we refer to our estimator as ``Two-Stage TMLE".   
 To the best of our knowledge, Two-Stage TMLE is the first semiparametric efficient estimator that adaptively adjusts for both individual-level missingness and for covariate imbalance in CRTs.
It can be applied to estimate a range of causal parameters % in settings with cluster-level exposures and individual-level outcomes subject to missing measurement. %, including both population and sample specific effects \citep{Rubin1990, Imbens2004, Imai2008, Balzer2016SATE}.
% It can also be applied to 
and under a range of CRT study designs, including differing randomization schemes (e.g., pair-matched or not) and approaches to participant follow-up within clusters (e.g., cross-sectional sampling or longitudinal follow-up). 

\section{Brief Review of CRT Methods}\label{Sec:Review}

We provide an overview  of existing CRT methods in Table~\ref{Tab:Review} and refer the reader to \cite{Benitez2021} for a detailed review.
A simple two-stage approach to account for the  dependence of participants within clusters is to aggregate the individual-level data to the cluster-level and then implement an effect estimator appropriate for independent data, such as a $t$-test. % \citep{HayesMoulton2009}.
Use of an unadjusted effect estimator in the second stage avoids modeling assumptions and the risk of over-fitting, but by ignoring covariate information is inefficient (e.g., \cite{Fisher1932, Gail1996, MarkBook, HayesMoulton2009,  Colantuoni2015, Turner2017Analysis, Murray2018, Murray2020}). 

 In contrast, mixed models and generalized estimating equations (GEE) %with cluster robust standard errors, 
 typically  adjust for a number of baseline individual- and cluster-level covariates, providing an opportunity to improve  precision of effect estimates %a regression of the individual-level outcome on the cluster-level treatment and on the individual- and cluster-level covariates 
 \citep{LairdWare82, LiangZeger86}. 
 However, neither address the need for a pre-specified approach to select the  adjustment variables that optimize efficiency, while preserving valid statistical inference. Further, both are susceptible to allowing the estimator choice define the effect measure that is estimated (e.g., GEE with a logistic link yields estimates of the conditional odds ratio)
 \citep{Hubbard2010}. 

To the best of our knowledge, only three methods generally allow for estimation of marginal effects in CRTs, while adjusting for individual- and cluster-level covariates. First, in the covariate adjusted residuals estimator (CARE), cluster-level outcomes are compared with those predicted from an individual-level regression of the outcome on individual- and cluster-level covariates, but not the cluster-level treatment \citep{Gail1996,HayesMoulton2009}. Second, augmented-GEE extends GEE for the marginal  effect %(i.e.,  only regression coefficients for the intercept and cluster-level treatment) 
by including an ``augmentation'' term, inspired by the efficient influence function \citep{Stephens2012, Stephens2013, Stephens2014}. Finally, hierarchical TMLE extends TMLE for estimation of marginal effects with cluster-based exposures  (\cite{Balzer2018Hierarchical,Benitez2021}; overview in the Supplementary Materials).

With regards to missingness, an unadjusted effect estimator requires the strongest identification assumption: there are no common causes of missingness and outcomes (i.e., the missing-completely-at-random, or MCAR, assumption holds) \citep{Rubin1976}.
%Little&Rubin02
The other methods rely on a weaker identification assumption; essentially, that the outcome distributions among persons for which the outcome is measured versus missing are exchangeable conditional on the treatment arm and some subset of measured covariates  
\citep{Turner2017Analysis, Hossain2017bin, Hossain2017cont}. 

Combining Augmented-GEE with inverse probability weighting yields a double robust estimator (``DR-GEE''); it is nearly unbiased if either the outcome regression or the measurement mechanism (i.e., the conditional probability of outcome measurement given the treatment arm and covariates) is correctly specified \citep{Prague2016}. To the best of our knowledge, DR-GEE’s methodology and computing code have been limited to adjustment for baseline variables only.  %\citep{Prague2016}.   
Hierarchical TMLE also offers the potential for integrated precision gains and double robust adjustment.
However, these extensions remain to be fully studied. %
Here, we, instead, develop and evaluate Two-Stage TMLE to (1) control for potentially differential missingness in each cluster separately, and  (2) adaptively adjust for covariate imbalance to improve efficiency when estimating the intervention effect. Before
doing so, we present our motivating example.

\section{Motivating Example}
The SEARCH Study was a pair-matched, pragmatic CRT of 32 communities each with 10,000 persons in rural Kenya and Uganda (ClinicalTrials.gov: NCT01864603) \citep{Havlir2019}. 
SEARCH was designed to evaluate the population-level effects of annual multi-disease testing and universal treatment for persons with HIV (intervention) versus baseline multi-disease testing and country-guided treatment (active control) on a range of outcomes including incident HIV, viral suppression % (a marker of effective HIV treatment) 
among persons  with HIV, %and a range of community health outcomes including
hypertension control, and incident tuberculosis (TB). 

As with the vast majority of CRTs, outcomes were not measured among all SEARCH participants and the MCAR assumption was unreasonable for many endpoints. 
Additionally, despite matching prior to randomization \citep{Balzer2015Adaptive}, covariate imbalance  
was expected; however, it was unclear \emph{a priori} which covariates to include in the adjustment set for optimal gains in efficiency.
To  reduce bias from missingness on individual-level outcomes and to maximize 
precision during effect estimation in the SEARCH Study, we developed Two-Stage TMLE.

\section{Two-Stage TMLE}
\label{sec2}

In many CRTs,  outcomes are assessed through longitudinal follow-up of a closed cohort of participants. In the SEARCH Study, for example, the primary outcome was the three-year cumulative incidence of HIV: the proportion of community residents ($\geq$ 15 years) who were HIV-uninfected at baseline and became infected with HIV over the three-year study. 
To assess the treatment effect on such endpoints, a cohort of participants who are at risk of the outcome is defined in each cluster. 
For each participant, let $W$ denote their baseline covariates, $M$ be their post-intervention covariates, 
$\Delta$ be an indicator of outcome measurement,
and  $Y$ be the outcome of interest. The  outcome $Y$ is  only observed when $\Delta=1$. We also observe cluster-level covariates $E^c$ and the randomly assigned cluster-level intervention $A^c$. Throughout, superscript $c$ will be used to distinguish cluster-level variables from individual-level variables. We denote the observed data structure for a participant as $O =\left (E^c, W, A^c,  M, \Delta, \Delta Y \right)$.
 In the SEARCH Study, for example, $E^c$ included baseline HIV prevalence and male circumcision coverage; $W$ included age, sex, marital status, occupation,  education, and mobility; $A^c$ was a community-level indicator of randomization to the intervention;  $M$ was interim HIV testing; 
$\Delta$ was an indicator of HIV testing at year 3, and $Y$ was an indicator of having a confirmed HIV-positive diagnosis at year 3 testing.

\begin{figure}[!p]
\centering\includegraphics[width=0.75\textwidth]{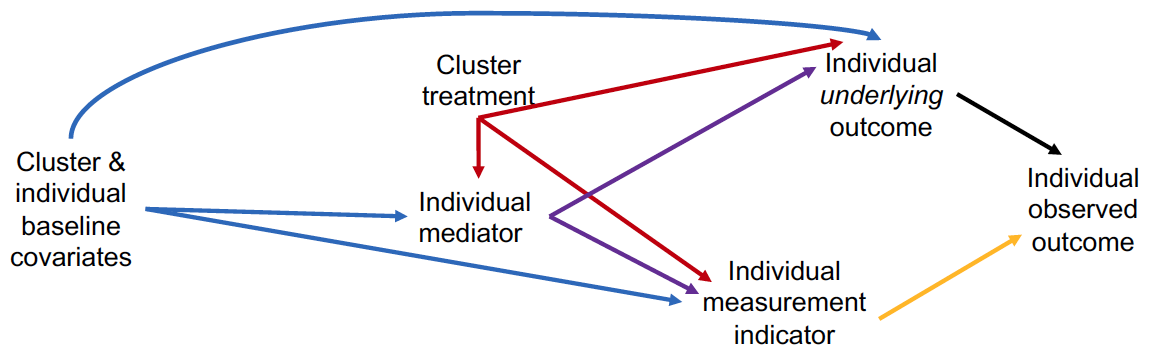}
\caption{Simplified causal graph to illustrate the challenges of a single stage approach to identifying effects defined by interventions on both the cluster-level treatment and individual-level measurement mechanism.}
\label{Fig:Complex}
\end{figure}

\begin{figure}[!p]
\centering\includegraphics[width=.6\textwidth]{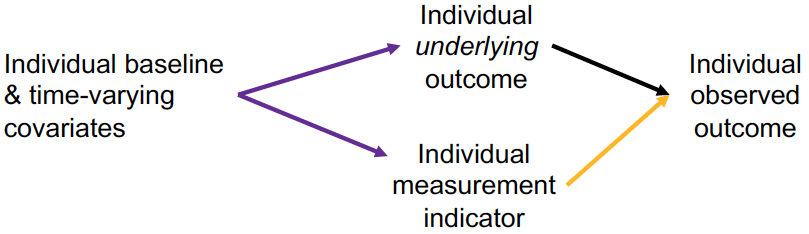}
\caption{Simplified causal graph to illustrate how stratifying on cluster in Stage 1 simplifies identification and estimation of the Stage 1 statistical parameter $Y^c$, corresponding to hypothetical intervention to ensure complete measurement. The Stage 1 estimates  $\hat{Y}^c$ are  used to evaluate the intervention effect in Stage 2.}
\label{Fig:Simplified}
\end{figure}

Recall our goal is to simultaneously control for differential missingness on individual-level outcomes, while estimating the effect of the cluster-level intervention with optimal precision. To do so, we consider an individual-level counterfactual outcome $Y(a^c, \delta)$, generated by hypothetical interventions  on the cluster-level treatment (i.e., to ``set'' $A^c=a^c$) and the measurement mechanism (i.e., to ``set'' $\Delta=\delta$). Identification of a corresponding causal parameter (e.g., $\mathbb{E}[Y(a^c,1)]$) is complicated by clustering and the missing data equivalent to time-dependent confounding (Figure~\ref{Fig:Complex}). % \citep{Robins1986,Robins2000, Robins1995, Robins2009}. 
Instead, in our novel Two-Stage approach, we separate control for missing  outcomes (Stage 1) from evaluation of the intervention effect (Stage 2). Specifically, in Stage 1, we fully stratify on each cluster, vastly simplifying identifiability and estimation for the missing data problem (Figure~\ref{Fig:Simplified}). Then in Stage 2, we use the estimates from Stage 1 to evaluate the intervention effect. 
Our approach  allows the missingness mechanism to vary by cluster, while avoiding specifying complex relationships between individual-level  ($W, M, \Delta$, $\Delta Y$) and cluster-level variables ($E^c, A^c)$.

\subsection{Stage 1: Identifying and Estimating Cluster-Specific Endpoints}

For the purposes of controlling for differential outcome measurement, we  consider each of the $N$ clusters separately in Stage 1. Since the cluster-level covariates and treatment $(E^c,A^c)$ are constant within each cluster, the Stage 1 observed data %for a participant
simplify to $O=(W,M,\Delta,\Delta Y)$ and the target causal parameter to  $\mathbb{E}[Y(\delta=1)]$.
Then if MCAR held in each cluster or, equivalently,  $Y(1) \independent \Delta$, this causal parameter could be identified as $\mathbb{E}(Y| \Delta=1)$ and consistently estimated  as the empirical mean among those measured within each cluster: $\hat{\mathbb{E}}(Y | \Delta=1)$. We relax this missing data assumption by allowing measurement to depend on the participant's baseline and time-varying characteristics $(W,M)$. Specifically, if 
$Y(1) \independent  \Delta \mid W, M$ and there is sufficient data support (i.e., the positivity assumption holds),  %\citep{Petersen2012}, 
our Stage 1 statistical estimand would be
\begin{equation}
\label{HIV}    
Y^c \equiv \mathbb{E}\big[ \mathbb{E}(Y \big| \Delta=1,W,M)
\end{equation}

Within each cluster separately, $Y^c$  could be estimated by a variety of algorithms, including inverse-weighting and G-computation \citep{Horvitz1952, Robins1986}.
We use TMLE for estimation the cluster-specific endpoint $Y^c$,
given its asymptotic properties and improved finite sample performance (e.g., \cite{MarkBook, Gruber2012}). 
% We refer readers  \cite{Schuler2017, Blakely2019} for an introduction.
Briefly, TMLE combines estimates of the outcome regression $\mathbb{E}(Y| \Delta=1,W,M)$
with those of the measurement mechanism $\mathbb{P}(\Delta=1|W,M)$. In doing so, TMLE achieves a number of desirable properties, including double robustness: a consistent estimate is attained if either the outcome regression or the measurement mechanism is consistently estimated. If both are consistently estimated at reasonable rates, TMLE will be efficient. In practice, we recommend implementing TMLE using Super Learner, an ensemble machine learning algorithm 
\citep{SuperLearner}. Step-by-step implementation for Eq.~\ref{HIV} is given in the Supplementary Materials. Since we are fully stratifying on cluster, an individual-level TMLE would be implemented $N$ times to obtain $N$ cluster-specific estimates: $\hat{Y}^c = \frac{1}{S^c} \sum_{j=1}^{S^c} \hat{\mathbb{E}}^*(Y \big| \Delta=1,W_j, M_j)$, where $j$ indexes the $S^c$ participants in a given cluster and $\hat{\mathbb{E}}^*(Y| \Delta=1,W_j, M_j)$ denotes the targeted prediction of the individual-level outcome for participant $j$ in that cluster.

%%%%%%%%%%%%%%%%%%%%%%%%%
\subsubsection{Stage 1 with Survival-Type Endpoints:}
\label{Sec:Surv}

When assessing  effects on time-to-event outcomes, participants are followed longitudinally until the occurrence of the event of interest or right-censoring. Examples of such endpoints 
in the SEARCH Study included the probability of treatment initiation %all-cause mortality, 
and the cumulative risk of HIV-associated TB or death due to illness. Our framework also accommodates survival-type endpoints. Specifically, to account for right-censoring in Stage 1, we would identify a different cluster-specific endpoint $Y^c$ and estimate it using the Kaplan-Meier estimator when censoring is non-differential or TMLE when censoring is differential \citep{Petersen2014ltmle,  Benkeser2019}.  

\subsubsection{Stage 1 with Endpoints Measured in a Cross-Sectional Design:} 
Other endpoints may be assessed using a cross-sectional design, where participants are measured at a single timepoint. In these settings,  we may  have missingness on  the characteristic defining the sub-population of interest as well as the outcome of interest. Consider, for example, population-level HIV viral suppression, defined as  
the proportion of all HIV-infected persons whose plasma HIV RNA level is suppressed below some limit: $\mathbb{P}(Suppressed \mid HIV+)$. 
Both baseline and time-varying factors  impact  HIV status and its measurement as well as  viral suppression and its measurement. 
To handle missingness on both the outcome (viral suppression) and the conditioning set (HIV-positivity), we redefine the outcome as the joint probability of being HIV-infected and suppressing viral replication, divided by HIV prevalence: $\mathbb{P}(Suppressed, HIV+) \div \mathbb{P}(HIV+)$. 
Again our Two-Stage approach can accommodate this ratio-type endpoint. Specifically, we would identify estimate (with TMLE) the numerator and denominator separately, and then take their ratio to estimate a new cluster-specific endpoint $Y^c$ in Stage 1 \citep{Balzer2020Supp}.

%%%%%%%%%%%%%
\subsection{Stage 2: Estimation of the Effect of the Cluster-level Intervention}
\label{Sec:VL}

Recall in Stage 1, we stratify on each cluster to identify and estimate a cluster-specific endpoint $Y^c$ that accounts for potentially differential measurement or censoring at the individual-level (e.g., Eq.~\ref{HIV}). Then in Stage 2, our goal is to use those estimates to evaluate the intervention effect with maximum precision. 
The Supplementary Materials provide a detailed discussion of several Stage 2 causal effects, which are all easily identified due to randomization of the cluster-level treatment $A^c$ and prior control for missingess  in Stage 1.
For estimation of these effects in Stage 2, the observed data  can be simplified to the cluster-level: $O^c = (E^c, W^c, A^c, \hat{Y}^c)$, where $E^c$ represents the baseline cluster-level covariates; $W^c$ denotes summary measures of the baseline individual-level covariates; $A^c$ is an indicator of randomization to the intervention arm, and $\hat{Y}^c$ is the estimated cluster-specific endpoint from Stage 1. 

With these data, a simple estimator of the treatment effect is the average outcome among intervention clusters $\hat{\mathbb{E}}(\hat{Y}^c| A^c=1)$ contrasted with the average  outcome among  control clusters $\hat{\mathbb{E}}(\hat{Y}^c| A^c=0)$.
Instead, to obtain a more efficient estimate of the intervention  effect (e.g., \cite{Moore2009, Rosenblum2010}), we implement a cluster-level TMLE  in Stage 2.  Briefly, an initial estimator of the cluster-level outcome regression $\mathbb{E}(\hat{Y}^c | A^c, E^c, W^c)$ is updated based on an estimate of the cluster-level propensity score $\mathbb{P}(A^c=1| E^c, W^c)$ to achieve a targeted estimator $\mathbb{E}^*(\hat{Y}^c | A^c, E^c, W^c)$. Targeted estimates of the expected outcomes under the intervention $\mathbb{E}^*(\hat{Y}^c | A^c=1, E^c, W^c)$ and under the control $\mathbb{E}^*(\hat{Y}^c | A^c=0, E^c, W^c)$ are generated for all clusters, averaged across clusters, and contrasted% to estimate the intervention effect
. Step-by-step implementation of TMLE in Stage 2 is given in the Supplementary Materials.

Unfortunately, it is nearly impossible to \emph{a priori}-specify the optimal estimators of the outcome regression  $\mathbb{E}(\hat{Y}^c | A^c, E^c, W^c)$ and propensity score $\mathbb{P}(A^c=1| E^c, W^c)$. Few clusters prohibit the use of Super Learner.
To avoid over-fitting while flexibly selecting the adjustment variables that maximize precision, we recommend using \emph{Adaptive Pre-specification} in Stage 2  \citep{Balzer2016DataAdapt}. The procedure data-adaptively selects from a pre-specified set the candidate adjustment variables and, thus, the TMLE that maximize empirical efficiency. Briefly, we pre-specify 
(1) candidate adjustment variables that are predictive of the outcome,
(2) a loss function corresponding to the squared influence curve for the TMLE of the target effect, and 
(3) a sample-splitting scheme; leave-one-out cross-validation is recommended. 
The candidate adjustment variables then define the set of candidate estimators for the outcome regression and the propensity score based on ``working'' generalized linear models (GLMs).  %\citep{Rosenblum2010}.
The procedure data-adaptively choses the combination of the outcome regression and propensity score GLMs (and thus the TMLE) with the lowest cross-validated variance estimate. Adaptive Pre-specification is an extension of Collaborative-TMLE using a cross-validation selector %(a.k.a., ``discrete Super Learner'') 
to maximize precision in  small trials. 

In CRTs with few clusters, we recommend limiting the candidate GLMs to a single adjustment covariate.
In CRTs with many clusters, we could also include candidate GLMs adjusting for multiple covariates and allow the procedure to data-adaptively determine the size of the adjustment set. 
%If, on a rare occasion, there are a large number of clusters, we could consider implementing Collaborative-TMLE with (non-discrete)  Super Learner using the squared influence curve as loss function.
Simulations mimicking the real data application (e.g., in sample size $N$ and expected within cluster dependence) can help inform these choices. 
Importantly, there is no guarantee that adjusting for more covariates  will improve empirical efficiency. 
For some CRTs,  none of the pre-specified covariates will improve precision over the unadjusted estimator. In this setting, 
the procedure will select 
the unadjusted effect estimator. Altogether with Adaptive Pre-specification, decisions about whether and how to adjust for optimal precision gains in Stage 2 are made with a rigorous procedure that does not compromise Type-I error.

%%%%%%%%%%%%%%%%%%%%%%%%%%

\subsection{Statistical Inference for Two-Stage TMLE}\label{Stage2inference}

Now we have a point estimate of the intervention effect and are ready to obtain statistical inference, which  occurs at the cluster-level. %For illustration, we focus an influence curve-based approach to estimate the variance of Two-Stage TMLE for the  $\psi(a^c)= \mathbb{E}[Y^c(a^c)]$. 
Under the following conditions, detailed in the Supplementary Materials, Two-Stage TMLE
  $\hat{\psi}$ is an asymptotically linear estimator of the intervention effect $\psi$, meaning that  $\hat{\psi}- \psi = \frac{1}{N}\sum_{i=1}^N IC_i + R_N$, 
where $IC_i$ is the cluster-level influence curve and $R_N= o_P(N^{-1/2})$ is the remainder term, going to zero in probability \citep{vanderVaart1998}: 
\begin{enumerate}
\item  Stage 1 estimation of the cluster-level outcomes $\hat{Y}^c$ provides negligible contribution to $R_N$;
\item   Stage 2 estimators of the cluster-level outcome regression and the cluster-level propensity score satisfy the usual regularity conditions (e.g., \cite{Moore2009}).
\end{enumerate}
The second condition is automatically satisfied when Adaptive Pre-specification is used to select among GLMs for the outcome regression $\mathbb{E}(\hat{Y}^c | A^c, E^c, W^c)$ and the known propensity score $\mathbb{P}(A^c=1 | E^c, W^c)$.
However, 
to satisfy the first condition we need 
(i) the Stage 1 estimators of the individual-level outcome regression and the individual-level measurement mechanism to converge to their targets at fast enough rates, 
(ii) the within cluster dependence to be weak enough that the Central Limit Theorem applies in cluster size $S_i^c$, and (iii) the cluster size is large relative to the total number of clusters (i.e., $N/min(S_i^c) \rightarrow 0$; Supplementary Materials). 

While these are asymptotic requirements, the Stage 1 conditions highlight the importance of having sufficiently sized clusters to support our missing data assumptions and to allow for flexible estimation of the individual-level outcome regression and measurement mechanism in Stage 1. Small cluster sizes may force us to rely on strong identifiability assumptions (e.g., MCAR within each cluster or only dependent on a single covariate) and strong estimation assumptions (e.g., the probability of being measured is accurately described by a main terms logistic regression).  Such assumptions may or may not be reasonable in a given application.
Larger cluster sizes, however, permit the use TMLE with Super Learner to flexibly adjust for the baseline and time-dependent covariates influencing outcomes and measurement in each cluster in Stage 1.
TMLE is also double robust, providing a consistent estimate of the cluster-specific endpoint $Y^c$ if either the individual-level outcome regression or measurement mechanism is consistently estimated.
Altogether, the requirements for valid statistical inference are not unique to our proposed approach; they apply to other two-stage estimators (e.g., a $t$-test on the mean outcome among those measured). In all cases, if the cluster-level endpoints $Y^c$ are estimated poorly in Stage 1, we are risk of  biased point estimates and misleading conclusions in Stage 2.   Simulations, below, explore the finite sample performance of Two-Stage TMLE under challenges commonly faced by CRTs: few clusters of modest size,  correlated outcomes within clusters, and differential outcome measurement.

%%%%%%%%%%%%%%%%%%%%%%%%%%%%%%%%%%%%%%%%
When the above conditions hold, the limit distribution of the standardized estimator is normal with mean 0 and variance given by the variance of its influence curve.  For the treatment-specific mean 
$\psi(a^c)=\mathbb{E}[\mathbb{E}(Y^c|A^c=a^c, E^c,W^c)]$, for example, the influence curve for Two-Stage TMLE  is approximated as 
$\hat{IC}(a^c) = \frac{\mathbb{I}(A^c=a^c)}{\hat{\mathbb{P}}(A^c=a^c | E^c,W^c)} [\hat{Y}^c -  \hat{\mathbb{E}}^*(\hat{Y}^c | A^c=a^c, E^c, W^c) ] + \hat{\mathbb{E}}^*(\hat{Y}^c | A^c=a^c, E^c, W^c) - \hat{\psi}(a^c)$.
We obtain a variance estimate with the sample variance of the estimated influence curve divided by the number of independent units $N$. Then using the Student's $t$-distribution with $N-2$ degrees of freedom as a finite sample approximation to the normal distribution
%\citep{Cornfield1978, Gail1996, HayesMoulton2009},
\citep{HayesMoulton2009}, we can  construct Wald-Type 95\% confidence intervals and conduct hypothesis testing. 
Additionally, through the Delta Method, we can derive the influence curve and variance estimator for the intervention effect on any scale of interest.  For the absolute effect $\psi(1) - \psi(0)$,  the estimated influence curve for TMLE would be  $\hat{IC}(1) -  \hat{IC}(0)$. For the relative effect $\psi(1) \div \psi(0)$,  we would apply the Delta method on the log-scale \citep{Moore2009}.  This approach to statistical inference also applies when the treatment is randomized  within matched pairs of clusters (Supplementary Materials).

\section{Simulation Study}\label{Sec:Sim}

We examine the finite sample performance of our proposed estimator using simulations to incorporate common CRT challenges, such as few randomized clusters and differential missingness.  Specifically, we focus on a setting with $N=30$ clusters and where within each cluster, the number of individual participants is sampled with equal probability from \{100, 150, 200\}. 
In these simulations, both baseline and post-baseline covariates impact  measurement of individual-level outcomes. 
Additional simulations and computing code are given in the Supplementary Materials.

\subsection{Data Generating Process} 
%%%%%%%%%%%%%%%%%%%%%%%%%%%%%%%%%%% 
 For each cluster $i=\{1,\ldots, N\}$, we independently generate the cluster-specific data as follows. First, three cluster-level, latent variables are independently generated as $U1^c \sim Unif(-1,1)$, $U2^c \sim Unif(-1, 1)$, and $U3^c\sim Norm(0,1)$.  Then, two individual-level covariates $(W1,W2)$ are drawn independently from normal distributions with cluster-specific means: $W1 \sim Norm(U1^c, 0.5)$ and $W2 \sim Norm(U2^c, 0.5)$. We set the observed cluster-level covariates $(E1^c, E2^c)$ as the empirical mean of their individual-level counterparts. The cluster-level intervention $A^c$ is randomly allocated within pairs of clusters matched on $U3^c$. %; therefore, $N/2$ clusters receive the intervention and $N/2$ the control. 

The individual-level mediator $M$ is generated as an indicator that $U_M\sim Unif(0,1)$ is less than the  $logit^{-1}\{-1  + 2 A^c + W1 + W2 + 0.2(1-A^c)( E1^c + E2^c) + 0.25U3^c\}$.
The underlying, individual-level outcome $Y$ is generated as an indicator that $U_Y\sim Unif(0,1)$ is less than $logit^{-1}(1  -2.5A^c + 4M + 0.5W1 + 0.5W2 + 0.2E1^c + 0.2 E2^c +  0.25U3^c)$. Finally,  individual-level  measurement  is generated as an indicator that $U_\Delta \sim Unif(0,1)$ is less than $A^c  logit^{-1}(3 -3M - 0.5W1 - 0.5W2) + (1-A^c)   logit^{-1}(-2 +3M + 0.5W1 + 0.5W2)$.
Thus, the measurement mechanism is highly differential by treatment arm with a mean of 70\% measured in the intervention arm and 43\% in the control arm. The observed outcomes $Y$ are set to missing for individuals with $\Delta=0$.

We also generate the counterfactual mediators and outcomes by setting the cluster-level treatment $A^c=a^c$ and preventing missingness (i.e., setting $\Delta=1$). The cluster-level, counterfactual outcome $Y^c(a^c)$ is the average of the individual-level, counterfactual outcomes within each cluster. We generate a population of 5000 clusters and calculate the true value of the treatment-specific, population means $\mathbb{E}[Y^c(a^c)]$ for $a^c=\{0,1\}$, their difference, and their ratio.

\subsection{Estimators Compared in the Simulation Study}

We compare a variety of estimators commonly implemented in CRTs. We consider 4 complete-case approaches, in which 
the data are subset to exclude participants with missing outcomes (i.e., those with $\Delta=0$): an unadjusted estimator, CARE, mixed models, and GEE. We also implement 2 approaches which use data on all participants:  
DR-GEE and our Two-Stage TMLE.

For the unadjusted approach, we 
first aggregate the individual-level outcomes $Y$ to the cluster-level $\hat{Y}^c$ by taking the empirical mean among those measured (i.e., $\Delta=1$) and then contrast the average  cluster-level outcomes $\hat{Y}^c$ by treatment arm $A^c$ with inference from the $t$-distribution. For CARE, we pool data across clusters and run  logistic  regression of the individual-level outcome $Y$ on the baseline covariates $(W1, W2, E1^c, E2^c)$; calculate the residuals by taking the difference between the cluster-level outcomes $\hat{Y}^c$ (the empirical mean among those measured) and those predicted from the previous regression, and finally use a $t$-test to compare the residuals by arm. 

In mixed models and  GEE, we again pool data across clusters and fit a log-linear regression of the individual-level outcome $Y$ on the cluster-level treatment $A^c$ and baseline covariates $(W1, W2, E1^c, E2^c)$. 
In DR-GEE, we also estimate the measurement mechanism with a  pooled logistic regression of $\Delta$ on $(W1, W2, E1^c, E2^c, A^c)$
 and  the augmentation terms with arm-specific log-linear regressions of $Y$ on \\ $(W1, W2, E1^c, E2^c)$. 
To account for  within cluster dependence, we include a random cluster-specific intercept in mixed models and use an independent working correlation matrix in the GEEs.
For mixed models, GEE, and DR-GEE, the default settings of \texttt{lme4}, \texttt{geepack}, and \texttt{CRTgeeDR} packages are used for standard error estimation, respectively. %%\citep{lme4, geepack, Prague2017}. 

For Two-Stage TMLE, 
  we first implement an individual-level TMLE within each cluster separately to estimate the cluster-specific endpoint $Y^ c \equiv \mathbb{E}\big[\mathbb{E}(Y \mid \Delta=1, W1, W2, M) \big]$. In these TMLEs, the outcome regression and the measurement mechanism are estimated using Super Learner to combine predictions from main terms logistic regression, generalized additive models, and the empirical mean. 
  In Stage 2, we  compare these cluster-specific estimates $\hat{Y}^c$ by treatment arm using a cluster-level TMLE with  Adaptive Pre-specification to select the optimal adjustment variables from $\{E1^c, E2^c, \emptyset \}$
  and with inference via the estimated influence curve (Section~\ref{Stage2inference}).

\subsection{Simulation Results}

The average coefficient of variation was 0.24 in the intervention arm and 0.17 in the control arm, reflecting expected levels of dependence between individual-level outcomes within clusters \citep{HayesMoulton2009}.
The true values of the %treatment-specific means $\psi(1)$ and $\psi(0)$ were 68.2\% and 77.3\%, respectively. The corresponding 
risk difference %$\mathbb{E}[Y^c(1)]-\mathbb{E}[Y^c(0)]$ 
and risk ratio %$\mathbb{E}[Y^c(1)]/\mathbb{E}[Y^c(0)]$ 
were -9.1\% and 0.88, respectively. 
For both  effects, Table~\ref{Tab:Sim2}   summarizes estimator performance when ``breaking the matches'' (i.e., ignoring the pair-matching scheme used for treatment randomization) and when preserving the matches. The exception is for DR-GEE, because to our knowledge, there does not yet exist an extension of DR-GEE for pair-matched CRTs. 
%For CARE, we only present results for the risk difference, because the recommended implementation for the relative effect yields an estimate of the geometric risk ratio, not the arithmetic risk ratio. For mixed models and the GEEs, only estimates for the risk ratio are presented, because the linear link is not recommended for binary outcomes. 

Focusing first on estimators of the risk difference (true value=-9.1\%), we see that $t$-test, which does not adjust for covariates, is highly biased, as expected given the differential measurement process. On average, it grossly overestimates the  effect by 22.9\% and attains confidence interval coverage of $<$1\%. By adjusting for covariates, CARE is less biased, but still overestimates the intervention effect by 12.7\% when breaking the matches and by 9.9\% when preserving the matches. The corresponding confidence interval coverages are much less than the nominal rate: 7.8\% and 40.4\%, respectively. In contrast, the bias of Two-Stage TMLE  is low ($<$1\%) and confidence interval coverage is good ($>$95\%). Also as predicted by theory \citep{ Balzer2015Adaptive}, more power is achieved when preserving (57.4\%) versus breaking the matches (52.8\%). 

Now focusing  on estimators of the risk ratio (true value=0.88), both mixed models and GEE overestimate the effect, preventing accurate inference. The confidence interval coverage  is 5.6-7.0\% for mixed models and 1.2-4.8\% for GEE. These estimators are expected to be unbiased when there are only baseline causes of missingness and the outcome regressions are correctly specified. Here, there are post-baseline causes of missingness, which are simultaneously mediators of the treatment-outcome relationship. %Standard analytic approaches are expected to fail in this setting \citep{Robins1986, Robins1995, Robins2009}. 
DR-GEE is expected to reduce bias due to missing outcomes by incorporating weights corresponding to the measurement mechanism, and this is seen when there are only baseline causes of missingness (Table 1 in Supplementary Materials). However, %to the best of our knowledge, 
extensions of DR-GEE to  handle post-baseline causes of missingness do not yet exist, and in the main  simulations,  DR-GEE attains $<$1\% confidence interval coverage for the risk ratio (Table~\ref{Tab:Sim2}). 

In contrast, Two-Stage TMLE is essentially unbiased and  achieves good confidence interval coverage ($>$95\%) for the risk ratio.  Again, more power is achieved when keeping (57.8\%) versus breaking the matches (52.6\%). In further simulation studies given in the Supplementary Materials, Two-Stage TMLE also performs well with fewer clusters  and maintains nominal Type-I error control when there is no effect. Finally, as shown in the Supplementary Materials, including the mediator $M$ in the adjustment set of the existing methods %CARE, mixed models, and the GEEs 
does not improve their performance.

\section{Application to the SEARCH Study}\label{Sec:App}

The results of the SEARCH Study have been previously published in \cite{Havlir2019}; here, we focus on  the efficiency gains from adjusting for covariate imbalance in Stage 2, \emph{after} adjusting for individual-level missingness in Stage 1. For select endpoints, we describe the estimator implementation and then compare point estimates and inference for the intervention effect when using TMLE with Adaptive Pre-specification versus the  unadjusted effect estimator in Stage 2. % We also compare breaking versus keeping the matched pairs used for randomization. 

As previously discussed, the primary outcome in the SEARCH Study was the three-year cumulative HIV incidence, measured in each community through a cohort of residents who were aged 15+ years and  HIV-uninfected at baseline. The pre-specified primary approach was Two-Stage TMLE to assess the  effect on the relative scale %, weighting communities equally, 
and keep the matched pairs.
In Stage 1, we estimated the community-specific, cumulative HIV incidence  $Y^c$ with TMLE adjusting for possibly differential capture of final HIV status. These individual-level TMLEs used Super Learner %, an ensemble  method 
to combine predictions from penalized regression, generalized additive models, main terms regression, and the empirical mean. In Stage 2, the intervention effect was estimated with a community-level TMLE using Adaptive Pre-specification to select the optimal adjustment set from baseline HIV prevalence, baseline male circumcision coverage, or nothing (unadjusted).

A similar approach was taken for all secondary endpoints, including  the incidence of HIV-associated TB or death due to illness, hypertension control %(systolic blood pressure $<$140 mmHg and diastolic blood pressure $<$90 mmHg) 
among adults (30+ years) with baseline hypertension, and population-level HIV viral suppression (HIV RNA$<$500 copies/mL). When assessing the impact on TB or death due to illness, we used the Kaplan-Meier method in Stage 1 to estimate the three-year risk in each community, separately; we censored at death due to other causes, outmigration, and study close.
When assessing the impacts on hypertension control and HIV viral suppression, we implemented individual-level TMLEs in Stage 1 to adjust for baseline and time-varying causes of missingness. 
For all secondary endpoints, we used a community-level TMLE with Adaptive Pre-specification to assess the intervention effect in Stage 2.  

As shown in Table~\ref{Tab:App}, the point estimates of the intervention effects are similar, but the precision gains from the Stage 2 approach are notable.  Here, ``efficiency'' is the variance of the unadjusted effect estimator breaking the matches divided by the variance of an alternative approach.
For the primary endpoint (HIV incidence), we see  precision gains when keeping versus breaking the matches; specifically, the unadjusted effect estimator is 3.1-times more efficient in the pair-matched analysis. As expected, TMLE with Adaptive Pre-specification keeping the matches is the most efficient approach and 4.6-times more efficient than the standard approach.

Similar results are seen for the incidence of HIV-associated TB and hypertension control (Table~\ref{Tab:App}). TMLE using Adaptive Pre-specification and keeping the matches is $\approx$2-times more efficient than the unadjusted effect estimator ignoring the matches.    
In contrast, minimal gains in efficiency are seen when evaluating the effect on HIV viral suppression. This is because the adaptive  approach used in TMLE  defaults to the unadjusted effect estimator  when adjustment does not improve precision. In this scenario, controlling for the baseline prevalence of viral suppression  or the proportion of youth (15-24 years) with HIV did not improve precision over the unadjusted effect estimator.  However, we note that assuming MCAR and relying on the unadjusted estimator in Stage 1 resulted in vast over-estimation of this endpoint in  
the intervention arm (85.2\% vs. 79.0\%) and control arm (75.8\% vs. 67.8\%) (Table 5 of the Supplementary Materials). %\citep{Balzer2020Supp}. 

\section{Discussion}
\label{sec4}

Cluster randomized trials (CRTs) are essential for assessing the effectiveness of interventions delivered to groups of individuals (e.g., clinics or communities). There have been notable advances in the  design and conduct of CRTs (e.g.,
%\citep{HayesMoulton2009, Crespi2016, Turner2017Design, Turner2017Analysis, Murray2018, Murray2020}. 
 \cite{ Turner2017Design, Turner2017Analysis, Murray2018, Murray2020}).
 However, substantial challenges remain and threaten the quality of evidence generated by CRTs. Regardless of best intentions, 
 most CRTs are prone to differential measurement of individual-level outcomes and to covariate imbalance.
In this paper, we proposed and evaluated a novel approach, Two-Stage TMLE, to address the dual challenges of bias due to missing individual-level outcomes and imprecision due to few randomized units (i.e., clusters). 
In Stage 1, an individual-level TMLE is implemented within each cluster separately to estimate a cluster-specific endpoint $Y^c$, which appropriately controls for missingness on participant outcomes. Fully stratifying on the cluster simplifies identification and allows the missingness mechanism to vary by cluster. 
In Stage 2, the treatment effect is estimated with a separate, cluster-level TMLE to compare the cluster-level endpoints $\hat{Y}^c$, estimated from Stage 1. Adaptive Pre-specification is used in Stage 2 to flexibly select from a pre-specified set the adjustment variables and the the TMLE that maximize precision \citep{Balzer2016DataAdapt}.  Statistical inference is based on the estimated influence curve and the Student's $t$-distribution. Finite sample simulations demonstrated the potential for Two-Stage to overcome the shortcomings of existing CRT methods, especially when there are post-baseline causes of missingness. Application to real data from the SEARCH Study demonstrated the precision gains attained through adaptive adjustment in Stage 2. 

To the best of our knowledge, Two-Stage TMLE is the first CRT estimator that simultaneously addresses bias due to individual-level missingness and improves efficiency through adaptive adjustment for covariate imbalance, in a fully pre-specified manner. The approach is applicable to a wide range of measurement schemes (e.g., single cross-sectional sample, repeated cross-sectional sampling, and longitudinal follow-up) and endpoint types (e.g., binary, continuous, time-to-event outcomes). The approach is also applicable to a wide range of causal parameters (e.g., population, conditional, and sample effects) and scales of inference (e.g., absolute or relative measures).
Additionally, Two-Stage TMLE should naturally generalize to hierarchical data settings with a non-randomized, cluster-level exposure. In such an observational setting, the cluster-level TMLE implemented in Stage 2 would focus on confounding control, as opposed to efficiency improvement. However, the asymptotic properties and finite sample performance of such an estimator remain an area of future work. 

Altogether, Two-Stage TMLE alleviates, but does not fully resolve, the challenges that arise from missing data in CRTs. Stage 1 uses TMLE with machine learning to flexibly adjust for baseline and time-dependent causes of missingness and, as a plug-in estimator, provides more stability under strong confounding or rare outcomes. However, adjustment for missingness in Stage 1 occurs within each cluster separately, limiting the breadth  and flexibility of adjustment when the cluster-specific outcome is rare or the cluster-specific sample size is small (e.g., in subgroup analyses). This highlights a challenge commonly occurring in finite samples: we must balance the strength of our  assumptions for identifiability (e.g., MCAR vs. MAR) and for estimation (e.g.,  parametric regressions vs. Super Learner) with limited data support. In CRTs with very small cluster sizes and highly differential measurement, a single stage TMLE is likely to be more appropriate. However, such an TMLE does yet not exist and is an area of future work.

\begin{table}[!p]
\caption{\emph{Description of CRT effect estimators as commonly implemented when outcomes are completely measured.}} 
\label{Tab:Review}
\centering
   \begin{tabular}{p{.12\linewidth}  p{0.84\linewidth}} 
   \hline
Unadjusted & Compare cluster-level outcomes by treatment arm; commonly implemented as a $t$-test. \\
CARE & At the cluster-level, compare observed outcomes with those predicted from a regression of the individual-level outcome on individual- and cluster-level covariates, but not the cluster-level treatment \citep{Gail1996,HayesMoulton2009}. \\
Mixed  Model & Point estimate and inference based the treatment coefficient in a regression of the individual-level outcome on the cluster-level treatment and individual- and cluster-level covariates; use random effects to account for dependence of individuals within a cluster \citep{LairdWare82}. \\
GEE & Point estimate and inference based the treatment coefficient in a regression of the individual-level outcome on the cluster-level treatment and individual- and cluster-level covariates; use a working correlation matrix to account for dependence of individuals within a cluster \citep{LiangZeger86}. \\
Augmented-GEE  & Modification to GEE for the marginal effect (i.e., GEE with only regression coefficients for the intercept and cluster-level treatment) by including an additional “augmentation” term for the outcome regression (i.e., the conditional expectation of the outcome given covariates and treatment) \citep{Stephens2012, Stephens2013, Stephens2014}.\\
%DR-GEE  & Modification to GEE for the marginal effect (i.e., GEE with only regression coefficients for the intercept and cluster-level treatment) by including an additional “augmentation” term for the outcome regression (the conditional expectation of the outcome, given covariates and treatment) and inverse-weighting to adjust for baseline predictors of missingness \citep{Prague2016}. \\
Hierarchical TMLE & %Compare targeted predictions of the outcomes under the intervention and control;
Modification of TMLE for cluster-based exposures; initial predictions of the outcome regression and propensity score (i.e., conditional probability of treatment given the covariates) are made by adaptively selecting between individual- or cluster-level specifications (\cite{Balzer2018Hierarchical, Benitez2021}; Supplementary Materials). \\
%Two-Stage TMLE & 1) In each cluster separately, run an individual-level TMLE to estimate the cluster-specific outcome, adjusting for baseline and time-dependent predictors of missingness. 2) Contrast the estimated outcomes by arm with a cluster-level TMLE. \\
\hline
\end{tabular}
\end{table}

\begin{table}[!p]
\caption{\emph{Over 500 simulated trials, the performance of CRT estimators when missingness depends on baseline and post-baseline variables. Results are shown when the target of inference is the risk difference (top 3 rows), when the target is the risk ratio (bottom 4 rows), when breaking the matches during analysis (left), and when preserving the matches during analysis (right).}} 
\label{Tab:Sim2}
\centering
\begin{tabular}{l | rrrrrr | rrrrrrr}
  \hline
 		& \multicolumn{6}{c}{BREAKING THE MATCHES} & \multicolumn{6}{c}{KEEPING THE MATCHES} \\ \hline
 & $\hat{pt}$ & bias & $\sigma$ & $\hat{\sigma}$ & CI  & power  &  $\hat{pt}$ & bias & $\sigma$ & $\hat{\sigma}$ & CI & power \\ 	
 \hline		& \multicolumn{12}{c}{FOR THE RISK DIFFERENCE (true value RD=-9.1\%)} \\ 
 t-test & -32.0 & -22.9 & 0.048 & 0.050 & 0.8 & 100.0 & -32.0 & -22.9 & 0.048 & 0.047 & 0.6 & 100.0 \\ 
  CARE & -21.8 & -12.7 & 0.037 & 0.037 & 7.8 & 100.0 & -19.0 & -9.9 & 0.049 & 0.040 & 40.4 & 98.0 \\ 
  TMLE & -9.8 & -0.7 & 0.038 & 0.046 & 98.8 & 52.8 & -9.9 & -0.8 & 0.037 & 0.043 & 96.6 & 57.4 \\ 	 \multicolumn{12}{c}{FOR THE RISK RATIO (true value RR=0.88)} \\ 
Mixed & 0.7 & -0.2 & 0.049 & 0.069 & 7.0 & 100.0 & 0.7 & -0.2 & 0.050 & 0.065 & 5.6 & 100.0 \\ 
  GEE & 0.7 & -0.2 & 0.049 & 0.056 & 4.8 & 100.0 & 0.7 & -0.2 & 0.055 & 0.036 & 1.2 & 99.8 \\ 
  DR-GEE & 0.7 & -0.2 & 0.049 & 0.054 & 0.2 & 100.0 &  &  &  &  &  &  \\ 
  TMLE & 0.9 & -0.0 & 0.051 & 0.063 & 98.4 & 52.6 & 0.9 & -0.0 & 0.051 & 0.058 & 96.8 & 57.8 \\  
  \hline
 \multicolumn{13}{l}{\footnotesize $\hat{pt}$: average point estimate (in \% for the RD)} \\
\multicolumn{13}{l}{\footnotesize bias: average deviation in the point estimates vs.  true effect (in \% for the RD)} \\
\multicolumn{13}{l}{\footnotesize$\sigma$: standard deviation of the point estimates (on log-scale for RR)} \\
 \multicolumn{13}{l}{\footnotesize $\hat{\sigma}$: average standard error estimate  (on log-scale for RR)} \\
 \multicolumn{13}{l}{\footnotesize CI: proportion of 95\% confidence intervals containing the true effect (in \%)} \\
  \multicolumn{13}{l}{\footnotesize power: proportion of trials correctly rejecting the false null hypothesis (in \%)}
\end{tabular}
\end{table}

\begin{table}[!p]
\caption{\emph{%Summary of effect measures for HIV incidence,  incidence of HIV-associated tuberculosis (TB),  hypertension control, and viral  suppression in the SEARCH Study.
For selected endpoints in the SEARCH Study, point estimates, 95\% confidence intervals, and efficiency comparisons  when estimating the intervention effect in Stage 2 with the unadjusted estimator and with TMLE using Adaptive Pre-specification. All approaches adjusted for individual-level missingness in Stage 1. }}
\label{Tab:App}
\centering
\begin{tabular}{ll | ll| ll}
  \hline
&  	& \multicolumn{2}{c}{\textbf{Breaking the Matches}} & \multicolumn{2}{c}{\textbf{Keeping the Matches}} \\ 
 & \textbf{Stage 2}	 & Effect (95\% CI) & Efficiency  & Effect (95\% CI) & Efficiency  \\ 
  \hline
 HIV Incidence & Unadjusted & 0.98 (0.66, 1.45) & 1 & 0.98 (0.78, 1.24) & 3.1 \\ 
&  TMLE  & 0.96 (0.73, 1.26) & 2.1 & 0.96 (0.8, 1.17) & 4.6 \\
TB Incidence & Unadjusted & 0.79 (0.64, 0.98) & 1 & 0.79 (0.69, 0.92) & 2.2 \\ 
  	& TMLE & 0.8 (0.67, 0.95) & 1.4 & 0.8 (0.69, 0.91) & 2.6 \\ 
Hypertension Control  & Unadjusted & 1.19 (1.1, 1.3) & 1 & 1.19 (1.11, 1.28) & 1.7 \\ 
  & TMLE & 1.18 (1.1, 1.26) & 1.6 & 1.19 (1.11, 1.27) & 1.8 \\ 
Viral Suppression & Unadjusted & 1.15 (1.11, 1.2) & 1 & 1.15 (1.11, 1.2) & 1 \\ 
  & TMLE & 1.16 (1.13, 1.2) & 1.1 & 1.15 (1.11, 1.2) & 1 \\ 
   \hline
\multicolumn{6}{l}{\footnotesize Efficiency: Variance estimate for the unadjusted effect estimator breaking the matches used for randomization, }\\
\multicolumn{6}{l}{\footnotesize divided by the variance  estimate of another approach (e.g., TMLE with Adaptive Pre-specification, keeping }\\
\multicolumn{6}{l}{\footnotesize the matches used for randomization).}
\end{tabular}
\end{table}

\section{Acknowledgments}

On behalf of the SEARCH Study, we thank the Ministry of Health of Uganda and of Kenya; our research  and administrative teams in San Francisco, Uganda, and Kenya; collaborators and advisory boards; and especially all the communities and participants involved. We also thank Dr. Ted Westling for his thoughtful feedback on this project.

This work was supported by the National Institutes of Health [grant numbers U01AI150510, U01AI099959, UM1AI068636, and R01AI074345]; by the President’s Emergency Plan for AIDS Relief; and by Gilead Sciences, which provided Truvada.

%%%%%%%%%%%%%%%%%%%%%%%%%%%%%%%

\section{SUPPLEMENTARY MATERIALS}

\begin{figure}[!ht] 
\centering\includegraphics[width=0.3\textwidth]{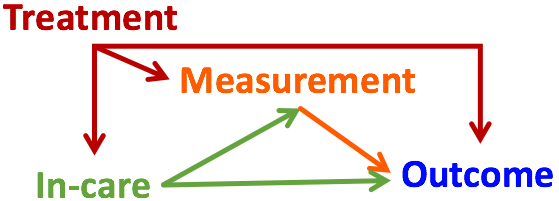}
\caption{Simplified causal graph to illustrate the challenges of adjustment for measurement impacted by the randomized treatment and post-baseline factors (here, being in care).}
\label{Fig:DAG}
\end{figure}

\subsection{Brief Overview of TMLE and of  Hierarchical TMLE}

The basic steps of targeted minimum loss-based estimation (TMLE) for a  point-treatment problem are as follows \citep{MarkBook}:
\begin{enumerate}
    \item  Estimating the outcome regression: the conditional expectation of the outcome given the intervention of interest and the adjustment covariates
    \item Estimating the propensity score: the conditional probability of the intervention given the adjustment covariates
    \item  Targeting the estimator of outcome regression with information in the estimated propensity score
    \item Obtaining a point estimate by averaging the targeted predictions of the outcome
    \item Obtaining statistical inference (i.e., Wald-Type 95\% confidence intervals) with the estimated influence curve
\end{enumerate}
We refer readers  \cite{Schuler2017} and \cite{ Blakely2019} for an introduction. Step-by-step implementation for statistical parameters corresponding to hypothetical interventions on the measurement process and for evaluating the treatment effect are given in Sections~\ref{Sec:Stage1} and~\ref{Sec:Stage2}, respectively.

Recently, \cite{Balzer2018Hierarchical} proposed and validated an extension of TMLE for estimation and inference for the effects of cluster-based  exposures in observational studies and randomized trials with complete outcome measurement (i.e., no missingness).  Briefly, this work explores the theoretical and finite sample performance of 3 different TMLEs: 
\begin{enumerate}
\item \emph{Cluster-level TMLE:} The cluster-level TMLE is implemented after the data are aggregated to the cluster-level. Initial estimation and targeting of the outcome regression are done at the cluster-level (i.e., with a cluster-level, propensity score estimate).
\item \emph{Hybrid-TMLE:} The Hybrid-TMLE is implemented using both individual-level  and cluster-level data. Initial estimation of the outcome regression is done at the individual-level, which naturally harnesses the pairing of individual-level outcomes and baseline covariates. Estimates from this individual-level outcome regression are aggregated to the cluster-level and targeted with a cluster-level, propensity score. 
\item \emph{Individual-level TMLE:} Point estimation for the individual-level TMLE follows a fully individual-level approach; statistical inference, however, respects the cluster as the independent unit. Initial estimation and targeting of the outcome regression are done at the individual-level (i.e., with an individual-level, propensity score estimate).
\end{enumerate}
These approaches are collectively known as ``Hierarchical TMLE" \citep{Balzer2018Hierarchical}.
Recently, \cite{Yang2021thesis} extended Adaptive Pre-specification to select between these TMLEs the one which maximizes the empirical efficiency. Additionally, 
\cite{Benitez2021} provide details on how weights can be applied to these TMLEs to estimate a variety of causal effects (e.g., effects at the individual-level and at the cluster-level; overview in subsection~\ref{Sec:Stage2Causal}). 

Hierarchical TMLE has  not yet been generalized to handle missingness on individual-level outcomes in CRTs. If there is no missingness and the individual-level outcome regression is fit within each cluster separately, then the Hybrid-TMLE can be considered to be a special case of Two-Stage TMLE, proposed here. Theoretically and in simulations mimicking the SEARCH Study but with complete outcome measurement, the Hybrid-TMLE dramatically increased efficiency and statistical power over the unadjusted effect estimator, while maintaining Type-I error control \citep{Balzer2018Hierarchical}. Therefore, in our Two-Stage approach, adjusting for individual-level covariates in Stage 1 is expected to increase the efficiency for effect estimation in Stage 2. When outcomes are completely measured, we can use Adaptive Pre-specification to select among the following TMLEs the one which maximizes empirical efficiency: (1) the cluster-level TMLE, (2) the Hybrid-TMLE where the individual-level outcome regressions are fit within each cluster separately, (3) the Hybrid-TMLE where the individual-level outcome regression is fit pooling over clusters, and (4) the fully individual-level TMLE. (We again note that approach \# 2 would be equivalent to Two-Stage TMLE when there is no missingness.)

Since participant outcomes are missing in over 90\% of CRTs \citep{Fiero2016}, we focus the remainder of the Supplementary Materials on Two-Stage TMLE, which simultaneously controls for differential outcome measurement and adjusts for covariate imbalance to reduce bias and improve efficiency in CRTs.

\subsection{Step-by-Step Implementation of TMLE in Stage 1}
\label{Sec:Stage1}

For demonstration, we focus on implementation of TMLE for the  cluster-specific endpoint \begin{equation}
\label{HIV}
Y^c \equiv \mathbb{E}\big[ \mathbb{E}(Y \big| \Delta=1,W,M) \big] 
\end{equation}
where $Y$ is the individual-level outcome, $\Delta$ is an indicator of measurement, $W$ are baseline individual-level covariates, and  $M$ are post-intervention individual-level covariates (i.e., mediators). We note that in settings with complex dependence, the adjustment set ($W,M)$ can be expanded to include the baseline and post-intervention covariates each participant's ``friends''.

To estimate Eq.~\ref{HIV} with TMLE,
we take the following steps \textbf{within each cluster $i=\{1, \ldots, N\}$, separately}. 
Throughout, $j=\{1, \ldots, S_i\}$ indexes the participants of cluster $i$. For ease of notation, we drop the superscript $c$ when denoting the cluster-size $S$ in the Supplementary Materials.
\begin{enumerate}
\item Among those with measured outcomes (i.e., $\Delta=1$), use Super Learner to flexibly model the relationship between the outcome $Y$ and adjustment variables $(W, M)$.
\item Use the output from \#1 to predict the outcome for all participants, regardless of their measurement status: $\hat{\mathbb{E}}(Y \mid \Delta=1, W_j, M_j)$ for $j=\{1,\ldots, S_i\}$.
\item Target these machine learning-based predictions with information in the estimated measurement mechanism $\hat{\mathbb{P}}(\Delta=1 \mid W, M)$, also fit with Super Learner. 
\begin{enumerate}
    \item Calculate the ``clever covariate" $\hat{H}_j= \frac{\mathbb{I}(\Delta_j=1)}{\hat{\mathbb{P}}(\Delta=1 \mid W_j, M_j)}$ for $j=\{1, \ldots, S_i\}$
    \item Run logistic regression of  outcome $Y$ on only the intercept, using the logit of the initial estimator $\hat{\mathbb{E}}(Y \mid \Delta=1, W, M)$ as offset (i.e., fixing its coefficient to 1) and the clever covariate $\hat{H}$ as weight.
    \item Denote the resulting intercept as $\hat{\epsilon}.$
\end{enumerate}

\item Obtain targeted predictions of the outcome for all participants, regardless of their measurement status: $\hat{\mathbb{E}}^*(Y \mid \Delta=1, W_j, M_j)$ for $j=\{1,\ldots, S_i\}$.
    \begin{enumerate}
    \item Add the estimated intercept to the logit of the initial estimates and transform back to the original scale (i.e., take the inverse-logit): $\hat{\mathbb{E}}^*(Y \mid \Delta=1, W_j, M_j)= logit^{-1}\big[\hat{\epsilon} + logit\{ \hat{\mathbb{E}}(Y \mid \Delta=1, W_j, M_j) \} \big] $ 
    \end{enumerate}
\item Average the targeted predictions to obtain an estimate of the cluster-specific endpoint  adjusted for missingness on individual-level outcomes: \[
\hat{Y}^c = \frac{1}{S} \sum_{j=1}^{S} \hat{\mathbb{E}}^*(Y \big| \Delta=1,W_j, M_j)
\]
\end{enumerate}
Because we are implementing TMLE in each cluster separately, we do not include the cluster-level covariates $E^c$ or treatment $A^c$ in the above estimation procedure. Updating on the logit-scale is recommended for binary and continuous individual-level outcomes; for details see \cite{Gruber2010}. 

This Stage 1 approach of identifying and then using TMLE to estimate a cluster-specific endpoint $Y^c$, which adjusts for differential outcome ascertainment, also applies to more complicated settings, including time-to-event outcomes with  differential censoring and when we have missingness  on both the  characteristic  defining  the population of  interest and on the  outcome  of  interest \citep{Petersen2014ltmle,  Benkeser2019, Balzer2020Supp}.
We refer the reader to Sections~\ref{Sec:Surv} and~\ref{Sec:VL} of the main text for an overview.

\subsection{Stage 2 Causal Parameters \& their Identification}
\label{Sec:Stage2Causal}

Recall our objective is to estimate the effect of the cluster-level intervention with optimal precision, after adjusting  for differential missingness on individual-level outcomes. Let $Y(a^c, 1)$ be the individual-level counterfactual outcome, generated by hypothetical interventions to set the cluster-level treatment $A^c=a^c$ and to ensure complete measurement of the individual-level outcomes (i.e., ``setting'' $\Delta=1$). As detailed in \cite{Benitez2021}, we can use these individual-level  counterfactuals to define a variety of cluster-level and individual-level effects in CRTs.
For example, we can define the cluster-level counterfactual outcome as the expectation of the individual-level counterfactual outcomes: 
 \begin{align}
Y^c(a^c) \equiv \mathbb{E}[Y(a^c, \delta=1)]
\label{Eq:ClusterCF}
\end{align}

 The Stage 2 causal parameter is then a summary measure of the distribution of the cluster-level counterfactuals $Y^c(a^c)$. A common target  is the population average treatment effect (PATE):  \begin{equation}
\mathbb{E}[ Y^c(1)]  - \mathbb{E}[  Y^c(0)]
 \end{equation}
Alternatively, we could be interested in the sample average treatment effect (SATE), which is the effect for the $N$ study clusters  \citep{Neyman1923, Rubin1990}, or in summary measures on the relative scale. In the SEARCH Study, for example, the primary analysis was for the  sample risk ratio for the $N=32$ trial communities: 
 \begin{equation}
 \label{Eq:SATE}
  \frac{ \frac{1}{N} \sum_{i=1}^N Y_i^c(1) }{ \frac{1}{N} \sum_{i=1}^NY_i^c(0) }  
  \end{equation}
  where $Y_i^c(a^c)$ was the counterfactual cumulative  incidence of HIV in community $i$.

We can  consider a wider range of causal parameters by combining  each summary measure with weights. %\citep{Balzer2018Hierarchical, Benitez2021}. 
Specifically, let $S_i$ be the size of cluster $i$, and consider a weighted-version of the treatment-specific sample mean: 
$1/N  \sum_i \alpha_i Y_i^c(a^c)$. Then setting $\alpha_i = \frac{S_i \times N}{\sum_i S_i}$ gives equal weight to participants, while setting $\alpha_i =1$ gives equal weight to  clusters \citep{Benitez2021}. 
 When there is an interaction between cluster size and the treatment, cluster size is said to be  ``informative'' \citep{Seaman2014},  and the resulting causal parameters will generally not be equivalent. 
 In all settings, the target effect should be pre-specified and be driven by research question. We refer the reader to \cite{Benitez2021} for a detailed discussion on target causal parameters in CRTs.
 
Since we have already controlled for missing outcomes in Stage 1, identification of the Stage 2 causal parameter is trivial. Specifically, the randomization assumption $(Y^c(a^c) \independent A^c)$ and the positivity assumption ($0<\mathbb{P}(A^c=1)<1$)  hold by design in CRTs. Therefore, we can identify PATE as 
 \begin{equation}
\mathbb{E}[ Y^c(1)]  - \mathbb{E}[  Y^c(0)] =  \mathbb{E}(Y^c | A^c=1) -  \mathbb{E}(Y^c | A^c=0)
 \end{equation}
 where $Y^c$ denotes the Stage 1 estimand, which appropriately adjusts for missingness on individual-level outcomes (e.g., Eq.~\ref{HIV} of the Supplementary Materials). This framework for specifying and identifying causal effects in Stage 2 also applies for more complicated Stage 1 endpoints, corresponding to different $Y^c$s. (See Sections~\ref{Sec:Surv} and~\ref{Sec:VL} of the main text.)
 
As repeatedly demonstrated (e.g., \cite{Gail1996,  Moore2009, Rosenblum2010, Colantuoni2015, Turner2017Analysis, Murray2020, Benkeser2020}), adjustment for baseline covariates can improve precision in randomized trials. Therefore, our statistical estimand corresponding to the treatment-specific, population mean $\mathbb{E}[Y^c(a^c)]$ is given by
 \begin{equation}
\psi(a^c)  \equiv  \mathbb{E}\big[ \mathbb{E}(Y^c | A^c=a^c \big| E^c, W^c) \big]
 \end{equation}
Likewise, our statistical estimand for the PATE is 
$\psi(1)  - \psi(0)$.
Of course, we can also take the ratio of $\psi(1)$ and $\psi(0)$ to obtain a relative effect. Identification of causal parameters for the corresponding sample and conditional effects is discussed in \cite{Balzer2016SATE}.

\subsection{Step-by-Step Implementation of the TMLE in Stage 2}
\label{Sec:Stage2}

Given estimates of the cluster-specific endpoints $\hat{Y}_i^c$ for $i=\{1,\ldots,N\}$ from Stage 1, we then implement a cluster-level TMLE to more efficiently estimate the intervention effect in Stage 2. For demonstration, we focus on TMLE for relative effect: $\psi(1)/\psi(0)$.

\begin{enumerate}
\item Obtain an initial estimate of the conditional expectation of the cluster-level outcome, given the cluster-level treatment and covariates: $\hat{\mathbb{E}}(\hat{Y}^c |A^c,E^c,W^c)$. We could, for example, fit a ``working'' regression  of the estimated outcome $\hat{Y^c}$ on an intercept with main terms for the cluster-level treatment $A^c$ and selected cluster-level covariates $(E^c,W^c)$ \citep{Moore2009, Rosenblum2010}. 

\item Use the output from \#1 to predict the outcome for all clusters under both the intervention and control conditions:  $\hat{\mathbb{E}}(\hat{Y}^c |A^c=1,E^c_i, W^c_i)$ and $\hat{\mathbb{E}}(\hat{Y}^c |A^c=0,E^c_i, W^c_i)$ for $i=\{1,\ldots, N\}$.

\item Target the initial predictions using information in the estimated propensity score $\hat{\mathbb{P}}(A^c =1 | E^c_i, W^c_i)$ for $i=\{1,\ldots, N\}$.
\begin{enumerate}
\item To estimate the cluster-level propensity score, we could again fit a ``working'' logistic regression  of the cluster-level treatment indicator $A^c$ on an intercept and selected cluster-level covariates $(E^c, W^c)$. 
\item Calculate the two-dimensional ``clever'' covariate: $\hat{H}1^c_i = \frac{\mathbb{I}(A^c_i=1)}{\hat{\mathbb{P}}(A^c=1 | E^c_i, W^c_i)}$ and $\hat{H}0^c_i = \frac{\mathbb{I}(A^c_i=0)}{\hat{\mathbb{P}}(A^c=0 | E^c_i, W^c_i)}$ for $i=\{1,\ldots, N\}$.
\item Run logistic regression of cluster-level outcome $\hat{Y}^c$ on the clever covariates $\hat{H}1^c$ and $\hat{H}0^c$, suppressing the intercept, and using the logit of the initial estimator $\hat{\mathbb{E}}(\hat{Y}^c |A^c,E^c, W^c)$ as offset (i.e., fixing its coefficient to 1). 
\item Denote the resulting coefficient estimates corresponding to $\hat{H}1^c$ and $\hat{H}0^c$ as $\hat{\epsilon}1^c$ and $\hat{\epsilon}0^c$, respectively. 
\end{enumerate}

\item Obtain targeted predictions of the outcome for all clusters under both the intervention and control conditions:
\begin{eqnarray*}
 \hat{\mathbb{E}}^*(\hat{Y}^c |A^c=1,E^c,W^c) &=& logit^{-1}\big[ logit\{\hat{\mathbb{E}}(\hat{Y}^c |A^c=1,E^c,W^c)\} + \hat{\epsilon}1^c/\hat{\mathbb{P}}(A^c=1 | E^c,W^c) \big] \\
 \hat{\mathbb{E}}^*(\hat{Y}^c |A^c=0,E^c,W^c) &=& logit^{-1}\big[ logit\{\hat{\mathbb{E}}(\hat{Y}^c |A^c=0,E^c,W^c)\} + \hat{\epsilon}0^c/\hat{\mathbb{P}}(A^c=0 | E^c,W^c) \big]
\end{eqnarray*}

\item Obtain a point estimate by dividing the average of the targeted predictions under the intervention condition by the average of the targeted predictions under the control condition:
\[
TMLE =  \frac{\hat{\psi}^*(1)}{\hat{\psi}^*(0)} = \frac{ \frac{1}{N} \sum_{i=1}^N \hat{\mathbb{E}}^*(\hat{Y}^c \mid A^c=1, E_i^c, W^c_i) }{\frac{1}{N} \sum_{i=1}^N \hat{\mathbb{E}}^*(\hat{Y}^c \mid A^c=0, E_i^c, W^c_i)}  
\]
\end{enumerate}
If the known propensity score is not estimated (e.g., $\mathbb{P}(A^c=1)= 0.5$ in two-armed CRTs with balanced allocation), then the targeting step can be skipped. As detailed in \cite{Moore2009}, using a two-dimensional clever covariate during updating (step 3) allows for simultaneous targeting of the treatment-specific means and effects on the additive, relative, and odds ratio scales. As detailed in \cite{Balzer2016SATE}, implementation to obtain a point estimate is identical for the population, conditional, and sample effects. 

To flexibly select among various estimators of the outcome regression and propensity score, we recommend using \emph{Adaptive Pre-specification}, as described in the main text and detailed in \cite{Balzer2016DataAdapt}.

\subsection{Asymptotic Linearity of Two-Stage TMLE}

Briefly, an estimator is asymptotically linear if the difference between the estimator and the estimand behaves (in first order) as an empirical average of a mean-zero and finite variance function, known as the influence curve, of the unit data \citep{Bickel1993, vanderVaartWellner1996, MarkBook}. An asymptotically linear estimator will be consistent and normally distributed in its limit. Therefore, the Central Limit Theorem can be applied to construct 95\% confidence intervals and test the null hypothesis.

Recall that in Stage 1, we first define the cluster-specific outcome $Y^c$. If all individual-level outcomes are completely measured, then  $Y^c$ could be defined as the expected individual-level outcome within each cluster: $\mathbb{E}[Y]$. If the individual-level outcomes are missing-completely-at-random (MCAR), then $Y^c$ could be defined as the expected individual-level outcome among those measured: $\mathbb{E}[Y| \Delta=1]$. Likewise, if measurement $\Delta$ depends on individual-level, baseline covariates  $W$, then $Y^c$ could be defined as the expected individual-level outcome given measurement and those covariates, standardized with respect to the covariate distribution: $\mathbb{E}\big[ \mathbb{E}(Y| \Delta=1, W) \big]$. Extensions to scenarios with post-baseline causes of missingness and/or right-censoring follow analogously. 

Next, we estimate the cluster-specific outcome $Y^c_i$ within each cluster $i=\{1,\ldots, N\}$, separately. When outcomes are completely measured ($Y^c=\mathbb{E}[Y]$) or are missing-completely-at-random ($Y^c=\mathbb{E}[Y|\Delta=1]$), a simple and intuitive estimator is the empirical mean outcome among those measured. 
When outcomes are missing-at-random within values of the adjustment variables (e.g., $Y^c=\mathbb{E}[\mathbb{E}(Y| \Delta=1, W)]$), we recommend using TMLE with Super Learner for estimation of the cluster-specific outcome. The empirical mean outcome (among those measured) can be considered a special case of TMLE where the adjustment set is empty: $W=\{\}$.

To emphasize how the Stage 1 estimator depends on the individual-level data within each cluster, let $P_{i}$ denote the true distribution of the individual-level data in cluster $i$. Likewise, let $P_{i,S_i}$ denote the targeted estimator of that distribution based on $S_i$ individuals in cluster $i$.  %
Then we can write the Stage 1 cluster-specific estimand as $Y^c_i \equiv \Phi^c(P_i)$ and the Stage 1  cluster-specific plug-in estimator as $\hat{Y}^c_i \equiv \Phi^c(P_{i,S_i})$.

The Stage 2 cluster-level effect estimator is, therefore, a function of $\Phi^c(P_{i,S_i})$, $i=\{1, \ldots, N\}$. Consider, for example, the treatment-specific mean %$\psi(a^c) = \frac{1}{N} \sum_{i=1}^N Y_i^c(a^c)$
$\mathbb{E}[Y^c(a^c)]$
as our Stage 2 target parameter. 
Then the cluster-level TMLE of the corresponding statistical estimand $\psi(a^c) = \mathbb{E}[\mathbb{E}(Y^c|a^c, E^c)]$ in Stage 2  would be
\[
\hat{\psi}(a^c) = \frac{1}{N} \sum_{i=1}^N \hat{\mathbb{E}}^*(\hat{Y}^c | A^c=a^c, E^c_i)
=
\frac{1}{N} \sum_{i=1}^N \hat{\mathbb{E}}^*\big(\Phi^c(P_{i,S_i})  | A^c=a^c, E^c_i\big)
\]
(For ease of notation, we use $E^c$ to represent both the cluster-level covariates and aggregates of individual-level covariates (i.e. $W^c$) in this sub-section.)
An unadjusted effect estimator in  Stage 2  can again be considered a special case of the cluster-level TMLE where the adjustment set is empty: $E^c=\{\}$.

Under the following conditions, Two-Stage TMLE will be asymptotically linear, meaning that 
\[
\hat{\psi}(a^c)- \psi(a^c) = \frac{1}{N}\sum_{i=1}^N D_i + R_N
\]
where $D_i$ represents the cluster-level influence curve and $R_N= o_P(N^{-1/2})$ is remainder term, going to zero in probability:
\begin{enumerate}
\item Stage 2 estimators of the cluster-level outcome regression and the cluster-level propensity score meet the usual regularity conditions, which are quite weak in a randomized trial (e.g., \cite{Moore2009, Rosenblum2010}).
\item  Deviations between the estimated cluster-level outcomes and the true cluster-level outcomes,  \\
$\frac{1}{\sqrt{N}} \sum_{i=1}^N \Phi^c(P_{i,S_i}) - \Phi^c(P_{i})$, provide a negligible contribution to the remainder term $R_N$.  
\end{enumerate}
The conditions on Stage 2 estimation are satisfied when estimating the known, cluster-level propensity score with a ``working'' logistic regression and when estimating the cluster-level outcome regression with another ``working'' parametric regression (e.g., \cite{Moore2009, Rosenblum2010}).
However, to the best of our knowledge, all previously existing Two-Stage estimators (e.g., a t-test on the cluster-level means) have  simply ignored the contribution from estimating the cluster-level outcome to $R_N$. 
Suppose, for example, our Stage 1 estimator is the average outcome within each cluster: $\Phi^c(P_{i,S_i})= \hat{\mathbb{E}}_{P_{i,S_i}}(Y | \Delta=1)$. (Such an estimator would only be appropriate when the individual-level outcomes are completely measured or are missing-completely-at-random.) Since the individual-level outcomes are not i.i.d. within each cluster, we need the following to hold for this estimator's contribution to $R_N$ to be essentially zero:  (1) the within cluster dependence is weak enough that the Central Limit Theorem applies in $S_i$, and (2) the smallest cluster is much larger than the total number of clusters (i.e., $N/min_i(S_i) \rightarrow 0$). 

When the Stage 1 estimator $\Phi^c(P_{i,S_i})$ is a TMLE of the Stage 1 estimand $\Phi^c(P_{i})$, 
the relevant component of the remainder term $R_N$ can be written as
\begin{equation}
\frac{1}{\sqrt{N}} \sum_{i=1}^N \bigg[
(P_{i,S_i} - P_{i} )D^*_{i,P_{i,S_i}} + R_i(  P_{i,S_i}, P_{i}) \bigg]
\label{Eq:Rn}
\end{equation}
where $D^*_{i,P_{i,S_i}}$ and $R_i(  P_{i,S_i}, P_{i})$ are the cluster $i$-specific efficient influence curve and remainder terms, respectively.
As before, we  need that the within cluster dependence is weak enough such that %the Central Limit Theorem applies
$(P_{i,S_i} - P_{i} ) D^*_{i,P_{i,S_i}} = 
O_P(S_i^{-1/2})$ % and  
%thus $1/\sqrt{N} \sum_i (P_{i,S_i} - P_{i} )D^*_{i,P_{i,S_i}} = o_P(min_i (S_i^{-1/2}))$. % In  words, we need enough independence across individuals within each cluster that the Central Limit Theorem in $S_i$ applies. 
and that the ratio of total number of clusters to the cluster-size goes to zero (i.e., $N/min_i(S_i) \rightarrow 0$). 
We note that when the cluster-size $S_i$ is substantially larger than $N$, we can weaken this independence assumption to allow for a slower rate of convergence. 
Additionally, we need that estimators of the individual-level outcome regression and the individual-level  missingness mechanism %are not ``too" adaptive and that they to 
converge to their targets at fast enough rates such that 
$R_i(  P_{i,S_i}, P_{i}) = o_P(S_i^{-1/2})$ \citep{MarkBook}. %, which must be satisfied by the product of the convergence rates for the outcome regression and missingness mechanism. (Recall we are using double robust TMLE in each cluster $i$ in Stage 1.) 
Implementing Super Learner with highly adaptive LASSO (HAL) \citep{Benkeser2016} or internal sample-splitting can help ensure these conditions hold in practice \citep{Chpt27, Diaz2019}.

\subsubsection{Inference for Pair-matched Trials} 
This approach to statistical inference also applies in CRTs where the treatment is randomized  within matched pairs of clusters. Briefly,  let $O^c_{k1}$ and $O^c_{k2}$ denote the observed data for the first and second cluster within matched pair $k$, respectively. To obtain statistical inference for the effect in a pair-matched setting, we  replace $\hat{D}(O^c)$ with the following paired version:
%\begin{align*}
 $\hat{D}_{paired}(O^c_{k1}, O^c_{k2}) = \frac{1}{2} \left[  \hat{D}(O^c_{k1}) +  \hat{D}(O^c_{k2}) \right]$
%\end{align*}
\citep{Balzer2016SATE}.
Our variance estimator is then given by the sample variance of the paired influence curve divided by the number of pairs ($N/2$), and we use the Student's $t$-distribution with $N/2 - 1$ degrees of freedom \citep{HayesMoulton2009}.
This could naturally be extended to matched triplets in a three-armed trial.

%%%%%%%%%%%%%%%%%%%%%%
\subsection{Additional Simulation Study with Baseline (only) Causes of Missingness}\label{Sec:Sim}

Here, we consider a simplified scenario where only baseline (but not post-baseline) covariates impact the measurement of individual-level outcomes.  As before, we focus on a setting with $N=30$ clusters and where within each cluster, the number of individual participants is sampled with equal probability from \{100, 150, 200\}. 

For each cluster $i=\{1,\ldots, N\}$, we independently generate the cluster-specific data as follows. First, one latent variable $U1^c$ is drawn uniformly from (1.75, 2.25) and two additional variables $(U2^c, U3^c)$ are drawn independently from a standard normal distribution.  Then, two individual-level covariates $(W1,W2)$ are generated by drawing from a normal distribution with means depending on the cluster-level latent factors: $W1 \sim Norm(U1^c, 1)$ and $W2 \sim Norm(U2^c, 1)$. We set the observed cluster-level covariates $(E1^c, E2^c)$ as the empirical mean of their individual-level counterparts. The intervention $A^c$ is randomly allocated within pairs of clusters matched on $U3^c$; therefore, $N/2$ clusters receive the intervention and $N/2$ the control.

The underlying, individual-level outcome $Y$ is generated as an indicator that  $U_Y$, drawn from a Uniform(0,1), is less than $logit^{-1}\{-4  + 0.15A^c + 0.15   A^c   W1 + 0.4 W1 + 0.2 W2 + 0.5  E1^cW1  + 0.3(E1^c + E2^c + U3^c)\}$.
Finally, we  incorporate individual-level missingness by generating $\Delta$ as an indicator that $U_{\Delta}$, drawn from a Uniform(0,1), is less than $logit^{-1}(4 -0.25 A^c  - 0.75 A^c W1 - 0.75 W1 - 0.1 W2  - 0.5 E1^c -0.1 E2^c  )$.
Thus, participants in the intervention arm $(A^c=1)$, and especially those with higher values of $W1$, are more likely to have the outcome and also be missing.  The observed outcomes $Y$ are set to be missing for individuals with $\Delta=0$. 

We also generate the counterfactual, individual-level outcomes $Y(1,1)$ and $Y(0,1)$ by setting the cluster-level treatment to $A^c=1$ and $A^c=0$, respectively, and preventing missingness by setting $\Delta=1$.  As before, the cluster-level counterfactual outcome is the empirical mean of the individual-level counterfactual outcomes within each cluster $Y^c(a^c) \equiv 1/S_i \sum_{i=1}^{S_i} Y_i(a^c, 1)$. The true values of the treatment-specific, population means  $\mathbb{E}[Y^c(a^c)]$ for $a^c=\{1,0\}$, their difference, and their ratio are calculated for a population of 5000 clusters.  
We compare the same estimators as the main simulation study. 

\subsubsection{Results from the Second Simulation Study}

In this simulation study, the average coefficient of variation was 0.27 in the intervention arm and 0.33 in the control, reflecting higher than expected levels of dependence within clusters \citep{HayesMoulton2009}.
 The true values of the treatment-specific  means were  $\mathbb{E}[Y^c(1)]$=47.4\% and $\mathbb{E}[Y^c(0)]$=39.6\%. The corresponding risk difference and risk ratio were 7.7\% and 1.20, respectively.
 For both effects, Table~\ref{Tab:Sim1}   illustrates estimator performance in this simplified setting.
 
Focusing first on estimating the risk difference (true value=7.7\%), we see that $t$-test, which fails to adjust for any covariates, is highly biased, as expected given the differential measurement process. On average, it grossly underestimates the intervention effect by 12.4\% and attains a confidence interval coverage of $<$20\%, much lower than the nominal rate of 95\%. By adjusting for covariates that influence measurement and underlying outcomes, CARE is less biased, but still underestimates the intervention effect by 2.8\% when breaking the matches and by 5.1\% when preserving the matches. The corresponding confidence interval coverages for CARE are  less than the nominal rate: 90.2\% and 41.2\%, respectively. In contrast, the bias of Two-Stage TMLE for the risk difference is negligible, and the confidence interval coverage is good ($>$95\%). As predicted by theory \citep{Balzer2015Adaptive}, higher power is achieved when preserving, as compared to breaking, the matches: 57.2\% versus 46.8\%, respectively.

Now focusing  on estimating the risk ratio (true value=1.2), we see that both mixed models and GEE overestimate the intervention effect. This bias is substantial enough to prevent accurate inference. The confidence interval coverage is 39.6\%-40.6\% for mixed models and 22.4\%-36.4\% for GEE. Lower coverage for GEE is likely due to underestimation of the standard errors ($\hat{\sigma} <\sigma$). While both mixed models and GEE are adjusting for the appropriate variables, both are relying on a misspecified regressions.

Theoretically, DR-GEE should reduce bias from GEE by incorporating estimates of the missingness mechanism. Indeed, DR-GEE exhibits lower bias, but still does not obtain valid inference (confidence interval coverage of 54\%). This again highlights the need for flexible (i.e., data-adaptive) estimators of the individual-level outcome regression and measurement mechanism. 
In contrast, Two-Stage TMLE for the risk ratio is essentially unbiased and confidence interval coverage is good ($>$95\%). Again, more power is achieved when preserving (59\%) versus breaking the matched (46.8\%).

\subsection{Main Simulation Study - Additional Results}

In the following Tables, we provide additional results from the main simulation study, where both baseline and post-baseline variables ($W,M$) impact individual-level measurement and outcomes.

In Table~\ref{Tab:Sim2MMMM}, we provide the results for the main simulation study when CARE, mixed models, GEE, and DR-GEE include the mediator $M$ in their adjustment set. As expected, forcing adjustment for a variable impacted by the intervention (but also confounds the measurement-outcome relationship) does not serve to eliminate bias due to missing individual-level outcomes.

To assess performance with fewer clusters, we repeated the main simulation study with $N=20$ clusters. The results are given in Table~\ref{Tab:MainSim20} and echo the main findings. Even with limited numbers of clusters, Two-Stage TMLE essentially eliminates bias due to differential outcome measurement and achieves nominal confidence interval coverage ($\geq$95\%). Existing estimators exhibit substantial bias and yield misleading inferences. (Here, the mediator $M$ is not included in the adjustment sets for CARE, mixed models, GEE, and DR-GEE.)

To assess Type-I error control for Two-Stage TMLE, we repeated the main simulation study when there was no treatment effect (RD=0; RR=1). 
The results are given in Table~\ref{Tab:Sim2Supp} and demonstrate for $N=\{20,30,50\}$ clusters, Two-Stage TMLE maintains nominal Type-I error control ($\leq$5\%).

\subsection{Additional Results from the SEARCH Study}

The full statistical analysis plan for the SEARCH Study is available at \cite{Balzer2018SAP}.
In Table~\ref{Tab:AppSupp}, we provide a comparison of results when using an unadjusted estimator in Stage 1 and Stage 2 versus Two-Stage TMLE when estimating population-level HIV viral suppression (the proportion of all persons with HIV who are suppressing viral replication $<$500 copiess/mL) in each arm and corresponding the intervention effect \citep{Balzer2020Supp}.

\subsection{Computing code}

All simulations were conducted in \texttt{R} (v4.0.3) using the \texttt{nbpMatching}, \texttt{lme4}, \texttt{geepack}, %\texttt{sandwich}, 
\texttt{CRTgeeDR} \texttt{ltmle}, and \texttt{SuperLearner} packages \citep{R, nbpMatching, lme4,  geepack, Prague2017, ltmlePackageb, SuperLearnerPackage}.
Computing code to reproduce the simulation study is available at {\url{https://github.com/LauraBalzer/TwoStageTMLE}}.
Computing code used to analyze the SEARCH Study data is available at \url{https://github.com/LauraBalzer/SEARCH_Analysis_Adults}.

\begin{table}[!p]
\caption{\emph{Over 500 simulated trials each with $N=30$ clusters, the performance of CRT estimators \textbf{when missingness is only impacted by baseline variables} (i.e., the supplemental simulation study). Results are shown when the target of inference is the risk difference  (top 3 rows), when the target is the risk ratio (bottom 4 rows), when breaking the matches during analysis (left), and when preserving the matches during analysis (right).}}
\label{Tab:Sim1}
\centering
\begin{tabular}{l | rrrrrr | rrrrrrr}
  \hline
 		& \multicolumn{6}{c}{BREAKING THE MATCHES} & \multicolumn{6}{c}{KEEPING THE MATCHES} \\ \hline
 & $\hat{pt}$ & bias & $\sigma$ & $\hat{\sigma}$ & CI  & power  &  $\hat{pt}$ & bias & $\sigma$ & $\hat{\sigma}$ & CI & power \\ 	
 \hline		& \multicolumn{12}{c}{FOR THE RISK DIFFERENCE (true value RD=7.7\%)} \\ 
 t-test & -4.6 & -12.4 & 0.040 & 0.043 & 19.4 & 15.8 & -4.6 & -12.4 & 0.040 & 0.039 & 18.2 & 22.6 \\ 
  CARE & 4.9 & -2.8 & 0.019 & 0.028 & 90.2 & 32.0 & 2.6 & -5.1 & 0.013 & 0.023 & 41.2 & 1.8 \\ 
  TMLE & 7.2 & -0.6 & 0.023 & 0.036 & 99.4 & 46.8 & 7.2 & -0.5 & 0.023 & 0.031 & 98.8 & 57.2 \\ 
	& \multicolumn{12}{c}{FOR THE RISK RATIO (true value RR=1.20)} \\ 
	Mixed & 1.7 & 0.5 & 0.148 & 0.144 & 40.6 & 90.4 & 1.7 & 0.5 & 0.142 & 0.144 & 39.6 & 92.0 \\ 
  GEE & 1.6 & 0.5 & 0.148 & 0.135 & 36.4 & 90.8 & 1.7 & 0.5 & 0.171 & 0.097 & 22.4 & 95.6 \\ 
  DR-GEE & 1.4 & 0.2 & 0.099 & 0.085 & 54.0 & 92.8 &  &  &  &  &  &  \\ 
  TMLE & 1.2 & -0.0 & 0.053 & 0.084 & 99.6 & 46.8 & 1.2 & -0.0 & 0.053 & 0.072 & 99.0 & 59.0 \\ 
   \hline
\multicolumn{13}{l}{\footnotesize $\hat{pt}$: average point estimate (in \% for the RD)} \\
\multicolumn{13}{l}{\footnotesize bias: average deviation in the point estimates vs.  true effect (in \% for the RD)} \\
\multicolumn{13}{l}{\footnotesize$\sigma$: standard deviation of the point estimates (on log-scale for RR)} \\
 \multicolumn{13}{l}{\footnotesize $\hat{\sigma}$: average standard error estimate  (on log-scale for RR)} \\
 \multicolumn{13}{l}{\footnotesize CI: proportion of 95\% confidence intervals containing the true effect (in \%)} \\
  \multicolumn{13}{l}{\footnotesize power: proportion of trials correctly rejecting the false null hypothesis (in \%)}
\end{tabular}
\end{table}

\begin{table}[!p]
\caption{\emph{Over 500 simulated trials each with $N=30$ clusters, the performance of CRT estimators \textbf{when missingness depends on baseline and post-baseline variables} (i.e., the main simulation study) and \textbf{the mediator $M$ is included in the adjustment set for CARE, mixed models, GEE, and DR-GEE}. Results are shown when the target of inference is the risk difference (top 3 rows), when the target is the risk ratio (bottom 4 rows), when breaking the matches during analysis (left), and when preserving the matches during analysis (right).}}
\label{Tab:Sim2MMMM}
\centering
\begin{tabular}{l | rrrrrr | rrrrrrr}
  \hline
 		& \multicolumn{6}{c}{BREAKING THE MATCHES} & \multicolumn{6}{c}{KEEPING THE MATCHES} \\ \hline
 & $\hat{pt}$ & bias & $\sigma$ & $\hat{\sigma}$ & CI  & power  &  $\hat{pt}$ & bias & $\sigma$ & $\hat{\sigma}$ & CI & power \\ 	
 \hline		& \multicolumn{12}{c}{FOR THE RISK DIFFERENCE (true value RD=-9.1\%)} \\
 t-test & -32.0 & -22.9 & 0.048 & 0.050 & 0.8 & 100.0 & -32.0 & -22.9 & 0.048 & 0.047 & 0.6 & 100.0 \\ 
  CARE & -17.7 & -8.6 & 0.031 & 0.028 & 17.0 & 100.0 & -15.4 & -6.4 & 0.040 & 0.033 & 56.0 & 98.4 \\ 
  TMLE & -9.8 & -0.7 & 0.038 & 0.046 & 98.8 & 52.8 & -9.9 & -0.8 & 0.037 & 0.043 & 96.6 & 57.4 \\ 
   & \multicolumn{12}{c}{FOR THE RISK RATIO (true value RR=0.88)} \\ 
   Mixed & 0.8 & -0.1 & 0.040 & 0.064 & 54.8 & 99.8 & 0.8 & -0.1 & 0.040 & 0.064 & 54.2 & 99.8 \\ 
  GEE & 0.8 & -0.1 & 0.040 & 0.043 & 17.4 & 100.0 & 0.8 & -0.1 & 0.047 & 0.033 & 3.4 & 99.8 \\ 
  DR-GEE & 0.7 & -0.2 & 0.054 & 0.064 & 0.0 & 100.0 &  &  &  &  &  &  \\ 
  TMLE & 0.9 & -0.0 & 0.051 & 0.063 & 98.4 & 52.6 & 0.9 & -0.0 & 0.051 & 0.058 & 96.8 & 57.8 \\ 
   \hline
 \multicolumn{13}{l}{\footnotesize $\hat{pt}$: average point estimate (in \% for the RD)} \\
\multicolumn{13}{l}{\footnotesize bias: average deviation in the point estimates vs.  true effect (in \% for the RD))} \\
\multicolumn{13}{l}{\footnotesize$\sigma$: standard deviation of the point estimates (on log-scale for RR)} \\
 \multicolumn{13}{l}{\footnotesize $\hat{\sigma}$: average standard error estimate  (on log-scale for RR)} \\
 \multicolumn{13}{l}{\footnotesize CI: proportion of 95\% confidence intervals containing the true effect (in \%)} \\
  \multicolumn{13}{l}{\footnotesize power: proportion of trials correctly rejecting the false null hypothesis (in \%)}
\end{tabular}
\end{table}

\begin{table}[!p]
\caption{\emph{Over 500 simulated trials, the performance of CRT estimators \textbf{when missingness depends on baseline and post-baseline variables} (i.e., the main simulation study) and \textbf{there are only $N=20$ clusters}. Results are shown when the target of inference is the risk difference (top 3 rows), when the target is the risk ratio (bottom 4 rows), when breaking the matches during analysis (left), and when preserving the matches during analysis (right).}}
\label{Tab:MainSim20}
\centering
\begin{tabular}{l | rrrrrr | rrrrrrr}
  \hline
 		& \multicolumn{6}{c}{BREAKING THE MATCHES} & \multicolumn{6}{c}{KEEPING THE MATCHES} \\ \hline
 & $\hat{pt}$ & bias & $\sigma$ & $\hat{\sigma}$ & CI  & power  &  $\hat{pt}$ & bias & $\sigma$ & $\hat{\sigma}$ & CI & power \\ 	
 \hline		& \multicolumn{12}{c}{FOR THE RISK DIFFERENCE (true value RD=-9.1\%)} \\ 
t-test & -32.4 & -23.4 & 0.059 & 0.061 & 7.4 & 100.0 & -32.4 & -23.4 & 0.059 & 0.058 & 7.2 & 100.0 \\ 
  CARE & -21.3 & -12.2 & 0.048 & 0.044 & 26.2 & 99.4 & -17.9 & -8.8 & 0.063 & 0.049 & 65.2 & 87.4 \\ 
  TMLE & -9.9 & -0.8 & 0.047 & 0.054 & 95.8 & 39.0 & -9.9 & -0.8 & 0.048 & 0.051 & 95.2 & 37.6 \\ 
  & \multicolumn{12}{c}{FOR THE RISK RATIO (true value RR=0.88)} \\ 
Mixed & 0.7 & -0.2 & 0.067 & 0.086 & 24.4 & 98.8 & 0.7 & -0.2 & 0.067 & 0.082 & 22.2 & 99.0 \\ 
  GEE & 0.7 & -0.2 & 0.067 & 0.069 & 17.0 & 98.4 & 0.7 & -0.2 & 0.076 & 0.046 & 5.4 & 99.2 \\ 
  DR-GEE & 0.7 & -0.2 & 0.064 & 0.064 & 2.0 & 99.4 &  &  &  &  &  &  \\ 
  TMLE & 0.9 & -0.0 & 0.064 & 0.074 & 96.2 & 40.6 & 0.9 & -0.0 & 0.065 & 0.069 & 95.2 & 38.4 \\ 
   \hline
 \multicolumn{13}{l}{\footnotesize $\hat{pt}$: average point estimate (in \% for the RD)} \\
\multicolumn{13}{l}{\footnotesize bias: average deviation in the point estimates vs.  true effect (in \% for the RD)} \\
\multicolumn{13}{l}{\footnotesize$\sigma$: standard deviation of the point estimates (on log-scale for RR)} \\
 \multicolumn{13}{l}{\footnotesize $\hat{\sigma}$: average standard error estimate  (on log-scale for RR)} \\
 \multicolumn{13}{l}{\footnotesize CI: proportion of 95\% confidence intervals containing the true effect (in \%)} \\
  \multicolumn{13}{l}{\footnotesize power: proportion of trials correctly rejecting the false null hypothesis (in \%)}
\end{tabular}
\end{table}

\begin{table}[!p]
\caption{\emph{Over 500 simulated trials, the performance of Two-Stage TMLE (only) \textbf{when missingness depends on baseline and post-baseline variables} (i.e., the main simulation study) and \textbf{there is no intervention effect} (i.e., under the null).  Results are shown for $N=\{20,30,50\}$ clusters when the target of inference is the risk difference (top), when the target is the risk ratio (bottom), when breaking the matches during analysis (left), and when preserving the matches during analysis (right).}}
\label{Tab:Sim2Supp}
\centering
\begin{tabular}{l | rrrrrr | rrrrrrr}
  \hline
 		& \multicolumn{6}{c}{BREAKING THE MATCHES} & \multicolumn{6}{c}{KEEPING THE MATCHES} \\ \hline
 & $\hat{pt}$ & bias & $\sigma$ & $\hat{\sigma}$ & CI  & $\alpha$  &  $\hat{pt}$ & bias & $\sigma$ & $\hat{\sigma}$ & CI & $\alpha$ \\ 	
  \hline	%	& \multicolumn{12}{c}{\textbf{UNDER THE NULL}} \\
 \		& \multicolumn{12}{c}{FOR THE RISK DIFFERENCE (true value RD=0\%)} \\ 
$N=20$ clusters  & -0.3 & -0.3 & 0.046 & 0.050 & 95.6 & 4.4 & -0.3 & -0.3 & 0.046 & 0.045 & 95.0 & 5.0 \\ 
$N=30$ clusters & -0.5 & -0.5 & 0.035 & 0.043 & 97.6 & 2.4 & -0.5 & -0.5 & 0.035 & 0.039 & 95.8 & 4.2 \\
  $N=50$ clusters & -0.6 & -0.6 & 0.026 & 0.034 & 98.4 & 1.6 & -0.6 & -0.6 & 0.026 & 0.031 & 97.6 & 2.4 \\ 
	& \multicolumn{12}{c}{FOR THE RISK RATIO (true value RR=1.0)} \\ 
$N=20$ clusters & 	1.0 & -0.0 & 0.065 & 0.070 & 95.8 & 4.2 & 1.0 & -0.0 & 0.065 & 0.064 & 95.0 & 5.0 \\ 
$N=30$ clusters & 1.0 & -0.0 & 0.049 & 0.060 & 97.6 & 2.4 & 1.0 & -0.0 & 0.050 & 0.055 & 95.8 & 4.2 \\ 
  $N=50$ clusters & 1.0 & -0.0 & 0.037 & 0.048 & 98.4 & 1.6 & 1.0 & -0.0 & 0.037 & 0.044 & 97.4 & 2.6 \\ 
   \hline
\multicolumn{13}{l}{\footnotesize $\hat{pt}$: average point estimate (in \% for the RD)} \\
\multicolumn{13}{l}{\footnotesize bias: average deviation between $\hat{pt}$ \& true effect (in \% for the RD)} \\
\multicolumn{13}{l}{\footnotesize$\sigma$: standard deviation of the point estimates (on log-scale for RR)} \\
 \multicolumn{13}{l}{\footnotesize $\hat{\sigma}$: average standard error estimate  (on log-scale for RR)} \\
 \multicolumn{13}{l}{\footnotesize CI: proportion of 95\% confidence intervals containing the true effect (in \%)} \\
  \multicolumn{13}{l}{\footnotesize $\alpha$: proportion of trials incorrectly rejecting the true null hypothesis (in \%)}
\end{tabular}
\end{table}

\begin{table}[!ht]
\caption{\emph{Summary of arm-specific and effect measures for population-level HIV viral  suppression in the SEARCH Study.  Point estimates and 95\% confidence intervals are provided when assuming MCAR in Stage 1 and using an unadjusted effect estimator in Stage 2 (``Unadjusted'') versus when using Two-Stage TMLE to control for missing individual-level outcomes and improve efficiency when estimating the intervention effect (``TMLE''), both when breaking the matches used for randomization and keeping the matches.}}
\label{Tab:AppSupp}
\centering
\begin{tabular}{l | llll }
  \hline
     &  & & Breaking matches & Keeping matches \\ 
  \textbf{Estimator}	 & Intervention (95\% CI) & Control (95\% CI) & Effect (95\% CI) & Effect (95\% CI)  \\ 
  \hline
Unadjusted & 85.2\% (83.5\%, 86.8\%) & 75.8\% (73.5\%, 78.2\%) & 1.12 (1.08, 1.16) & 1.12 (1.09, 1.16) \\ 
  TMLE & 79\% (77.1\%, 80.8\%) & 67.8\% (66.2\%, 69.5\%) & 1.16 (1.13, 1.2) & 1.15 (1.11, 1.2) \\
   \hline
\end{tabular}
\end{table}

\newpage
\bibliography{Bibliography}

\end{document}